\documentclass[fleqn,usenatbib]{mnras}

\usepackage{newtxtext,newtxmath}

\usepackage[T1]{fontenc}

\DeclareRobustCommand{\VAN}[3]{#2}
\let\VANthebibliography\thebibliography
\def\thebibliography{\DeclareRobustCommand{\VAN}[3]{##3}\VANthebibliography}

\usepackage{graphicx}	
\usepackage{amsmath}	
\usepackage{fancyhdr}
\usepackage{pdflscape}

\title[The metallicity of M dwarfs $-$ II]{Calibrating the metallicity of M dwarfs in wide physical binaries with F-, G-, and K-primaries - II: Carbon, oxygen, and odd-Z iron-peak abundances of the primary stars}

\author[C. Duque-Arribas et al.]{
C. Duque-Arribas,$^{1}$\thanks{E-mail: chrduque@ucm.es}
H. M. Tabernero,$^{1}$
D. Montes$^{1}$
and J.\,A. Caballero$^{2}$
\\
$^{1}$ Departamento de F{\'i}sica de la Tierra y Astrof{\'i}sica \& IPARCOS-UCM (Instituto de F\'{i}sica de Part\'{i}culas y del Cosmos de la UCM), Facultad de Ciencias\\ \phantom{$^1$ }F{\'i}sicas, Universidad Complutense de Madrid, 28040 Madrid, Spain\\
$^{2}$ Centro de Astrobiolog\'{i}a (CSIC-INTA), camino bajo del Castillo s/n, 28691 Villanueva de la Ca\~{n}ada, Madrid, Spain
}

\date{Accepted 2024 January 4. Received 2023 December 07; in original form 2023 June 05}

\pubyear{2024}

\begin{document}
\label{firstpage}
\pagerange{\pageref{firstpage}--\pageref{lastpage}}
\maketitle

\begin{abstract}
Detailed chemical composition of stars is of prime interest for a range of topics in modern stellar astrophysics, such as the chemical evolution of the Galaxy or the formation, composition, and structure of exoplanets. In this work, we derive the C and O abundances and update Sc, V, Mn, and Co abundances considering hyperfine structure effects (HFS) and correcting for non-local thermodynamical equilibrium (NLTE) for a sample of 196 late-F, G-, and early-K stars with wide resolved M-dwarf companions. We accomplished this by employing the equivalent width (\textit{EW}) method and high-resolution spectroscopic data. Furthermore, we investigated the distributions of [X/Fe] ratios and [C/O] as a function of metallicity ([Fe/H]) and kinematic population. The observed trends are consistent with previous findings reported in the literature. Additionally, we searched for confirmed exoplanets around our primary stars in the literature and found 24 exoplanets in 17 systems, while none of the M-dwarf companions in our sample presented confirmed exoplanets.
In conclusion, our study provides homogeneous abundances from high-resolution spectra for a large sample of FGK primary stars, paving the way for further research on stellar abundances of the M secondaries and exoplanetary science.
\end{abstract}




\defcitealias{Montes2018MNRAS.479.1332M}{Paper I}
\defcitealias{Battistini2015A&A...577A...9B}{BB15}
\defcitealias{Delgado-Mena2021A&A...655A..99D}{DM21}

\section{Introduction}

\thispagestyle{fancy}
\fancyhead{} 
\fancyfoot{}
\fancyhead[L]{MNRAS (2024)\\ \small{https://doi.org/10.1093/mnras/stae076}}
\fancyhead[R]{IPARCOS-UCM-24-003\\ }
\renewcommand{\headrulewidth}{0pt}

Cool, low-mass M-type dwarf stars stand as widely interesting objects in astrophysics. Being the most abundant main-sequence stars in the Milky Way \citep{Henry2006AJ....132.2360H,Winters2015AJ....149....5W,Reyle2021A&A...650A.201R}, and having main-sequence lifetimes far longer than the present age of the Universe \citep{Adams1997RvMP...69..337A}, M dwarfs rank as excellent objects to study the structure and chemical evolution of the Galaxy \citep{Bahcall1980ApJS...44...73B,Reid1997PASP..109..559R,Chabrier2003PASP..115..763C,Ferguson2017ApJ...843..141F}. Additionally, M dwarfs constitute relevant targets in the search for exoplanets, since their faintness and lower masses facilitate the detection of low-mass exoplanets in their habitable zone \citep{Tarter2007AsBio...7...30T, Kopparapu2014ApJ...787L..29K, Shields2016PhR...663....1S}. As a result, new spectrographs have been developed specifically optimized for conducting exoplanet searches around M dwarfs, such as CARMENES \citep{Quirrenbach2020SPIE11447E..3CQ,Ribas2023A&A...670A.139R}.

These studies could be increased with prior knowledge of stellar metallicity and abundances. However, this duty is often really challenging in the case of M dwarfs, whose spectra are complex and notoriously difficult to model due to the presence of prominent molecular features, in comparison to those of solar-type stars \citep[e.g.][]{Allard2000ApJ...540.1005A,Passegger2018A&A...615A...6P,Marfil2021A&A...656A.162M}. For this reason, several studies have tried photometric calibrations of metallicity for M dwarfs \citep[see][and references therein]{DuqueArribas2023ApJ...944..106D}. Other studies have investigated M-dwarf metallicities using wide physical binary systems formed by a F-, G-, or K-primary star and a M-dwarf companion \citep[e.g.][]{Woolf2006PASP..118..218W,Bean2006ApJ...652.1604B,Rojas-Ayala2010ApJ...720L.113R,Terrien2012ApJ...747L..38T,Mann2013AJ....145...52M,Mann2014AJ....147..160M,Newton2014AJ....147...20N,Montes2018MNRAS.479.1332M,Ishikawa2020PASJ...72..102I,Souto2020ApJ...890..133S,Souto2022ApJ...927..123S}. These binary systems provide an excellent opportunity to test not only the metallicities of the M dwarfs with those for their warmer primaries, but also the chemical abundances for individual atomic species. This is the second item of a series of papers eventually aimed at calibrating the metallicity of M dwarfs in wide physical binary systems with solar-type stars. However, here we still focus mostly on the primary stars, especially on the carbon, oxygen, and odd-Z iron-peak abundances. As a result, a calibration of the metallicity of the M dwarfs is not presented here yet, but we will do so in the next item of the series.

Oxygen and carbon are, after H and He, the most abundant chemical elements in the Universe. Carbon plays a role in dust formation processes in the interstellar medium \citep{Leger1984A&A...137L...5L,Weingartner2001ApJ...548..296W,Draine2007ApJ...657..810D}, contributes significantly to the stellar interior and atmospheric opacity \citep{Burrows1997ApJ...491..856B,Baraffe1998A&A...337..403B,Meynet2002A&A...390..561M}, and it is an essential element for life as we know it, being important in searching for biomarkers in habitable exoplanets \citep{DesMarais2002AsBio...2..153D,Scalo2007AsBio...7...85S,Seager2010ARA&A..48..631S}. Furthermore, the oxygen abundance is used to map the chemical enrichment and star formation history of stellar populations in the Milky Way and to infer the metallicities from the H\,II regions \citep{Tolstoy2009ARA&A..47..371T,Moustakas2010ApJS..190..233M}. The studies of C and O abundances have become more and more popular in recent years in the context of determining the composition of terrestrial exoplanets. Theoretical models predict that carbon-to-oxygen and magnesium-to-silicon ratios (C/O and Mg/Si) may provide information about the structure and composition of planets; the C/O ratio controls the distribution of Si among carbide and oxide species, while Mg/Si determines the silicate mineralogy \citep{Larimer1975GeCoA..39..389L,Bond2010ApJ...715.1050B,SuarezAndres2018A&A...614A..84S}. Recent studies suggest that elemental abundance ratios may differ when studied in host stars and in planetary atmospheres: some ratios such as Mg/Si and Fe/Si show similar values, but larger differences are found for C/O due to the dependence of the distance for volatile elements \citep{Carter-Bond2012ApJ...760...44C, Marboeuf2014A&A...570A..36M, Thiabaud2015A&A...580A..30T}.

On the other hand, iron-peak elements, i.e. elements in the periodic table from Sc to Ge, are synthesised in thermonuclear explosions of supernovae and in Si-burning during explosive burning of core-collapse supernovae \citep{Kobayashi2016Natur.540..205K}. However, the stellar yields of these elements are under debate \citep{Kobayashi2020ApJ...900..179K}.
In particular, the accurate determination of the abundances for odd-Z iron peak elements (Sc, V, Mn, and Co) require non-local thermodynamic equilibrium modelling of spectral lines taking into account the hyperfine structure splitting. These elements are relevant in several astrophysical analysis. For instance, manganese is fundamental in stellar population and nucleosynthesis studies to constrain the physics of SNe Ia \citep{Seitenzahl2013A&A...559L...5S}. Cobalt is an interesting element in regard to the galactic chemical evolution, but there are disagreements about the overall abundance trend of Co in the halo and disc and about its nucleosynthesis production \citep{Bergemann2010MNRAS.401.1334B}. Additionally, scandium holds significance in comprehending Am and Fm stars, which exhibit an overabundance of iron-peak elements but a deficiency in scandium and calcium.

In the first paper of this series (\citealt{Montes2018MNRAS.479.1332M}, hereafter referred to as \citetalias{Montes2018MNRAS.479.1332M}), we established a sample of 192 wide physically-bound systems, derived precise stellar atmospheric parameters ($T_\text{eff}$, $\log g$, $\xi$, and chemical abundances for 13 atomic species) for the primary stars using the equivalent width (\textit{EW}) method and high-resolution spectra, under the local thermodynamic equilibrium (LTE) assumption, and performed a kinematic analysis, classifying the stars in different Galactic populations and stellar kinematic groups. In this second paper, we update the abundances of scandium (Sc), vanadium (V), manganese (Mn) and cobalt (Co) taking into account hyperfine structure (HFS) and non-local thermodynamic equilibrium (NLTE) effects, and determine new abundances for carbon (C) and oxygen~(O).
The LTE approximation simplifies energy distribution through particle collisions, but it becomes less accurate close to the stellar surface, where the radiation field deviates from being local, isotropic, and Planckian \citep{Steenbock1984A&A...130..319S,Rutten1988ASSL..138..185R,Thevenin1999ApJ...521..753T}. Additionally, interactions between electron and nuclear spins of species with non-zero nuclear spin cause energy states to split, resulting in multiple components in the absorption line of the corresponding transition \citep{Abt1952ApJ...115..199A,Kurucz1993PhST...47..110K,Jofre2017A&A...601A..38J,Heiter2021A&A...645A.106H}.

This paper is organised as follow. In Sect.~\ref{Sect_analysis} we describe the sample, line list and abundance analysis, indicating our solar abundances compared with previous results. Sect.~\ref{Sect_results} reports and discusses the results for C, O, Sc, V, Mn, and Co. Finally, Sect.~\ref{Sect_conclusions} outlines our findings and presents an outlook for the future items of this series.

\section{Analysis}
\label{Sect_analysis}

\subsection{Stellar sample}

\begin{figure}
    \centering
    \includegraphics[width=\columnwidth]{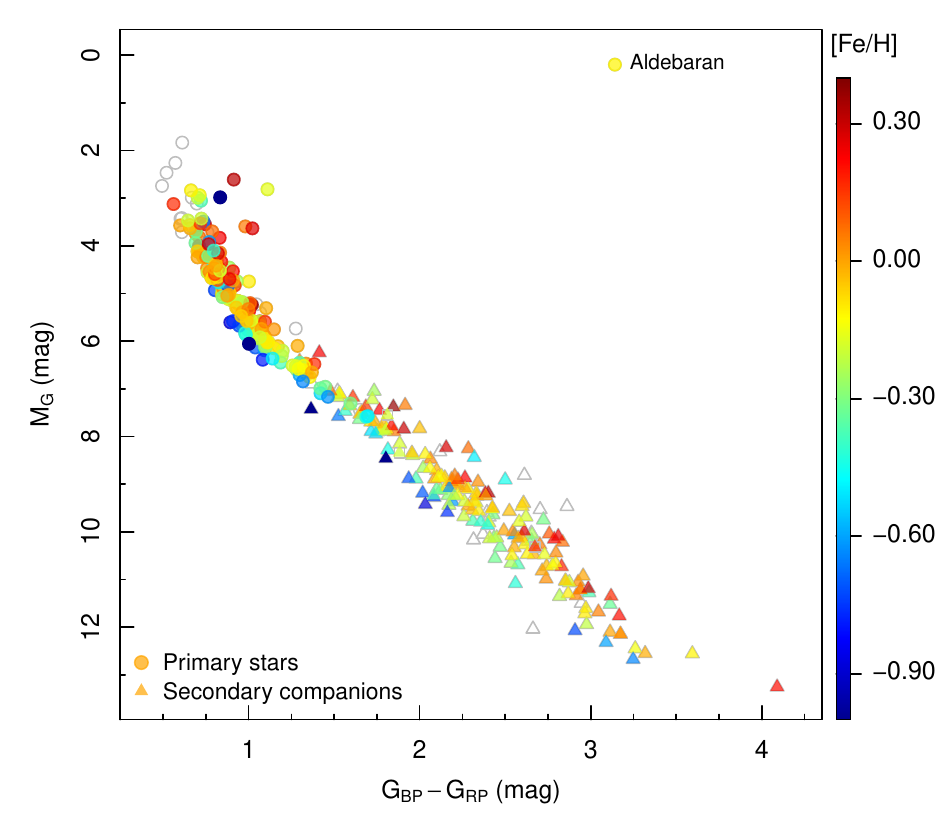}
    \caption{Colour-magnitude diagram of the primary and secondary stars in our sample using \textit{Gaia} DR3 data, colour-coded by primaries' [Fe/H]. Open data symbols are stars in systems whose primary is without derived stellar parameters.}
    \label{fig:colour_mag}
\end{figure}

The sample used in this work was presented in \citetalias{Montes2018MNRAS.479.1332M} and consists of 192 binary systems made of late-F, G-, or early-K primaries and late-K or M dwarf companion candidates. We carried out observations with the HERMES spectrograph \citep{Raskin2011A&A...526A..69R} at the 1.2\,m Mercator Telescope at the Observatorio del Roque de los Muchachos (La Palma, Spain) and obtained high-resolution spectra for 196 FGK-type stars. These spectra were analyzed with the automatic code \textsc{StePar}\footnote{\url{https://github.com/hmtabernero/StePar}} \citep{Tabernero2019A&A...628A.131T}, which relies on the equivalent width method, to derive precise stellar atmospheric parameters (effective temperature $T_{\rm eff}$, surface gravity $\log{g}$, microturbulence velocity $\xi$ and iron abundance [Fe/H]) for 179 stars. However, we were not able to determine the stellar parameters of the rest stars due to factors such as double-line spectroscopic binaries, fast-rotating stars, too hot stars ($T_\text{eff}>6700$\,K) for not having enough iron lines, and too cool stars ($T_\text{eff}<4500$\,K) with too many overlapping iron lines. We re-analised the proper motions and parallaxes of these systems using \textit{Gaia} DR3 \citep{GaiaCollaboration2022arXiv220800211G}, confirming the results of \citetalias{Montes2018MNRAS.479.1332M}. Therefore, we derived reliable spectroscopic stellar parameters for 174 primaries and 5 companions. For additional details on the basic properties of the studied systems and stellar parameters, please see Appendix B in \citetalias{Montes2018MNRAS.479.1332M}. In Fig.~\ref{fig:colour_mag} we show a colour-magnitude diagram of the primary and secondary stars in our sample.

\subsection{Abundances, HFS effects, and NLTE corrections}


\begin{figure*}
    \begin{tabular}{cc}
      \includegraphics[width=0.99\columnwidth]{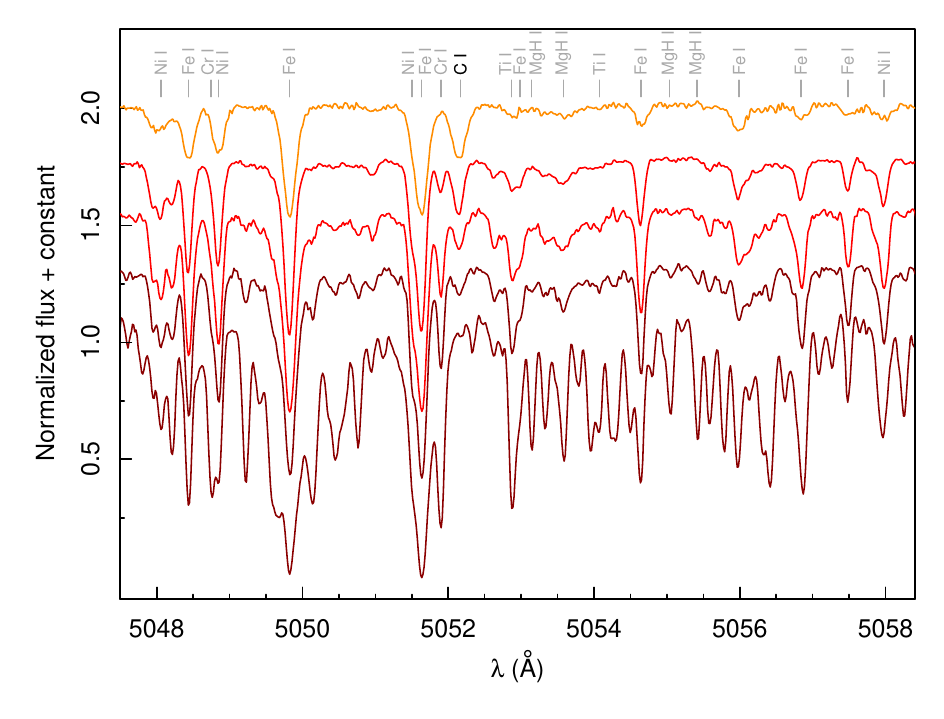} &
      \includegraphics[width=0.99\columnwidth]{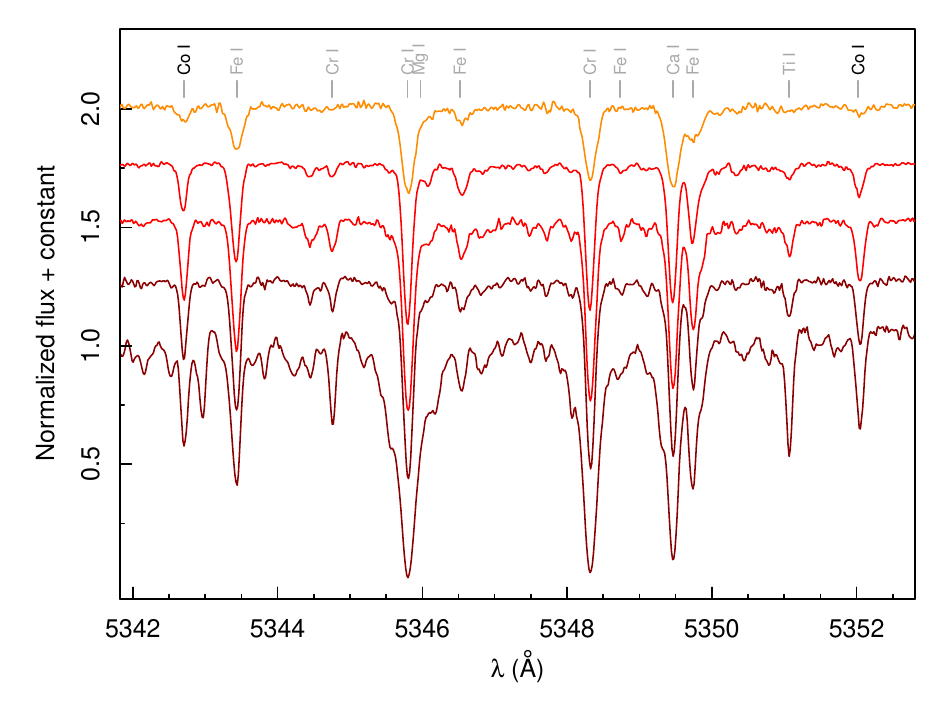} \\
      \includegraphics[width=0.99\columnwidth]{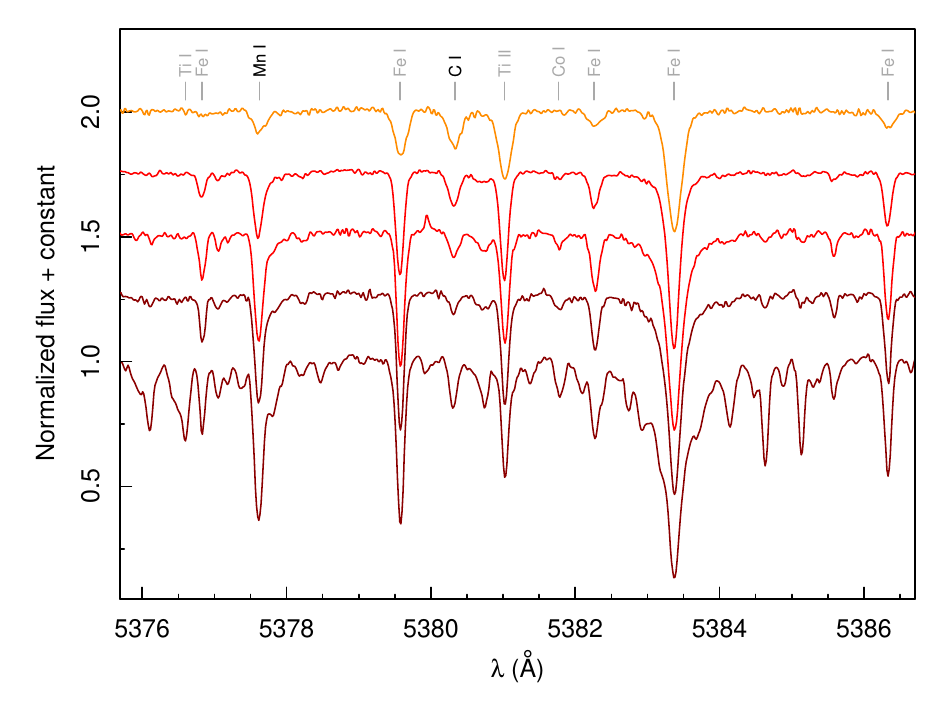} &
      \includegraphics[width=0.99\columnwidth]{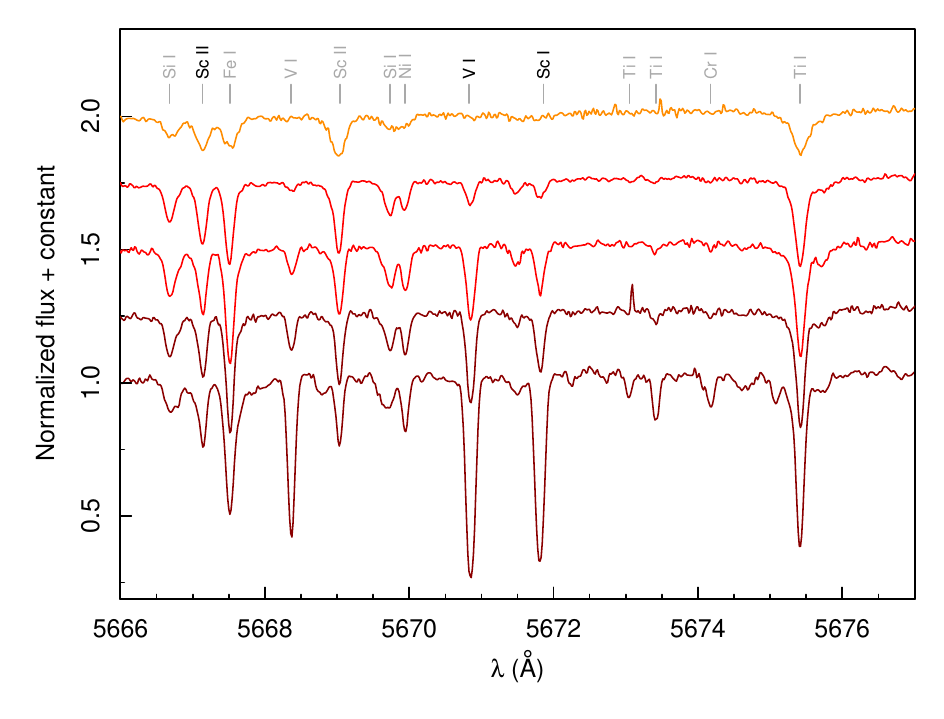} \\
      \includegraphics[width=0.99\columnwidth]{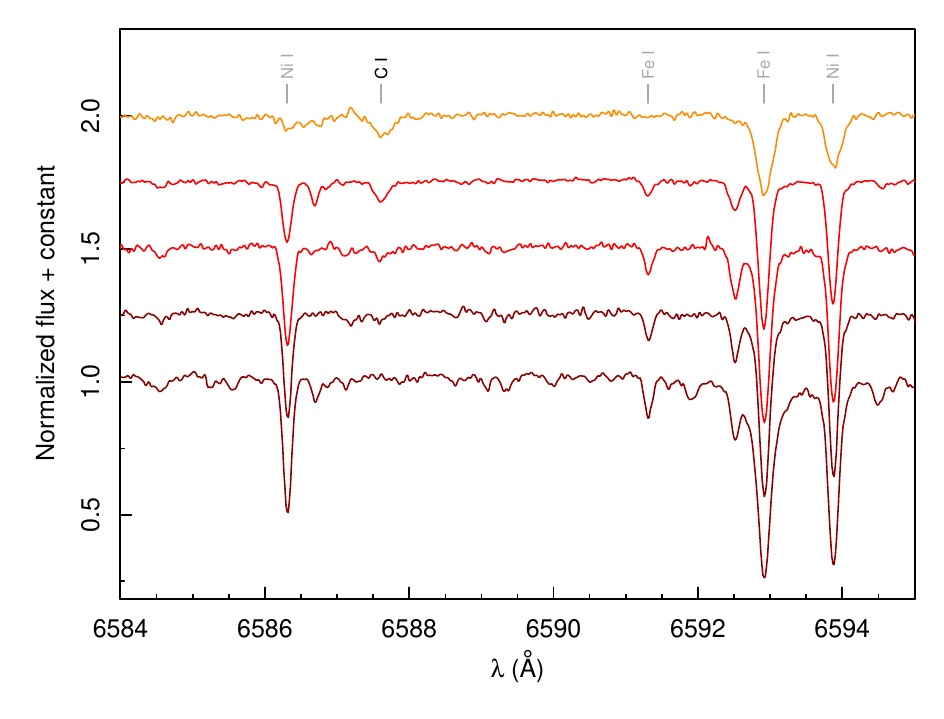} &
      \includegraphics[width=0.99\columnwidth]{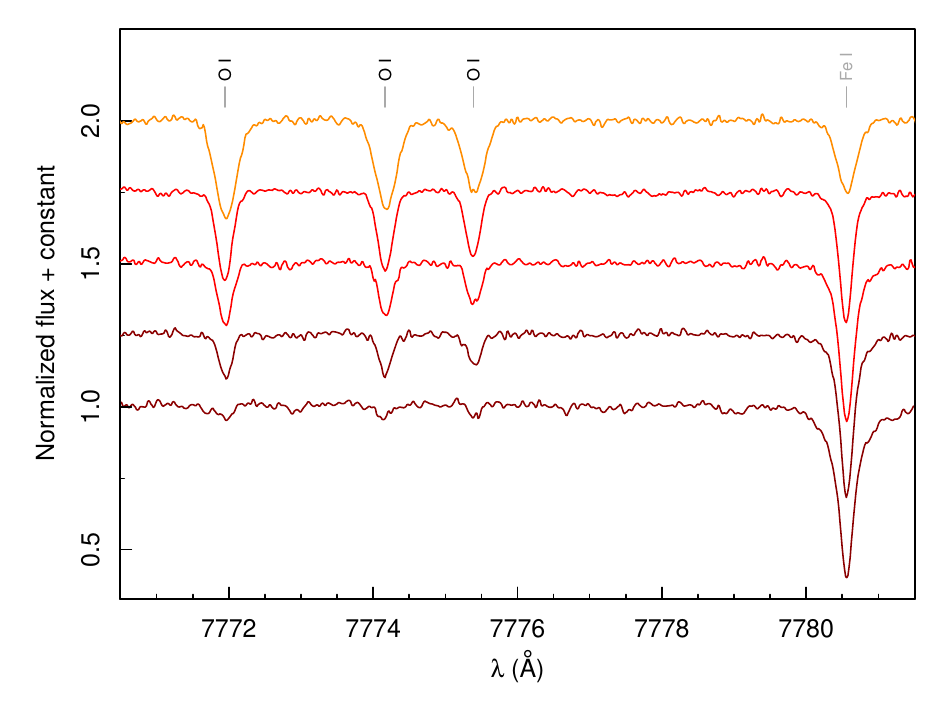} \\  
    \end{tabular}

    \includegraphics[width=0.5\textwidth]{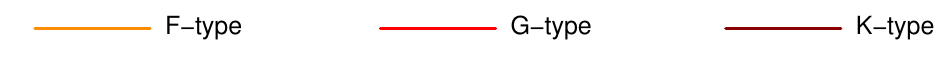}
    
    \caption{High-resolution spectra of five representative primaries from our sample of solar metallicity and different spectral types (\textit{from top to bottom}): HD 35638 (F5\,V), HD 43587 (G0\,V), HD 82939 (G5\,V), BD+68 1345 A (K0\,V), and HD 6660 A (K4\,V). Each panel shows zoomed ranges near some of the lines investigated.}
    \label{fig:spectra_CO}
\end{figure*}

To derive the carbon and oxygen abundances, we use the same line list as \cite{Tabernero2021A&A...646A.158T}, that is, \ion{C}{I} lines at $5052.17$, $5380.34$, and $6587.61$\,{\AA} and the \ion{O}{I} infrared triplet lines at $7771.9$, $7774.2$ and $7775.4$\,{\AA}. These \ion{C}{I} lines can be perfectly described in LTE. On the other hand, the \ion{O}{I} triplet, typically strong in solar-type stars and located in a spectral region free from other blending lines, is significantly affected by NLTE \citep{Kiselman1993A&A...275..269K, Asplund2009ARA&A..47..481A, Amarsi2015MNRAS.454L..11A}. In Fig.~\ref{fig:spectra_CO} we show these \ion{C}{I} and \ion{O}{I} spectral lines in five representative primaries, all of them with solar metallicity but covering different spectral types. We can see that the lines become very weak for the cooler stars, and the \ion{C}{I} $\lambda$5380\,{{\AA}} line may present an unknown blend for $T_\text{eff}<5200$\,K \citep{Delgado-Mena2021A&A...655A..99D}.

The line list used for the Sc, V, Mn, and Co abundances is the same as in \citetalias{Montes2018MNRAS.479.1332M}.
Table \ref{tab:lines} lists the atomic data for the lines analysed, retrieved from  the Vienna Atomic Line Database (VALD3, \citealt{Ryabchikova2015PhyS...90e4005R}).

\begin{table}
    \centering
    \scriptsize
    \caption{Atomic parameters for the lines used with the corresponding solar abundances in LTE, NLTE corrections, and number of HFS components.}
    \begin{tabular}{lcccccc} \hline
    \noalign{\smallskip}
         Element & $\lambda$ & $\log(gf)$ & $\chi_l$ & $A(\text{X})_\odot^\textsc{lte}$ & NLTE corr. & \# HFS  \\
         & ({\AA}) & & (eV) & & & comp. \\
    \noalign{\smallskip}
    \hline
    \noalign{\smallskip}
         \ion{C}{I} & $5052.17$ & $-1.304$ & $7.68$ & $8.359$ & $\cdots$ & $\cdots$ \\
         \ion{C}{I} & $5380.34$ & $-1.615$ & $7.68$ & $8.429$ & $\cdots$ & $\cdots$ \\
         \ion{C}{I} & $6587.61$ & $-1.021$ & $8.54$ & $8.327$ & $\cdots$ & $\cdots$ \\
         
         \ion{O}{I} & $7771.944$ & $0.369$ & $9.15$ & $8.779$ & $-0.081$ & $\cdots$ \\
         \ion{O}{I} & $7774.166$ & $0.223$ & $9.15$ & $8.789$ & $-0.070$ & $\cdots$ \\
         \ion{O}{I} & $7775.388$ & $0.002$ & $9.15$ & $8.769$ & $-0.052$ & $\cdots$ \\
         
         \ion{Sc}{I} & $5520.519$ & $0.290$ & $1.865$ & $3.321$ & $\cdots$ & 4 \\
         \ion{Sc}{I} & $5671.860$ & $0.500$ & $1.4478$ & $3.037$ & $\cdots$ & 21 \\
         \ion{Sc}{II} & $5526.789$ & $-0.010$ & $1.7682$ & $3.207$ & $\cdots$ & 20 \\
         \ion{Sc}{II} & $5657.886$ & $-0.540$ & $1.5070$ & $3.265$ & $\cdots$ & 13 \\
         \ion{Sc}{II} & $5667.135$ & $-1.210$ & $1.5004$ & $3.248$ & $\cdots$ & 7 \\
         \ion{Sc}{II} & $5684.190$ & $-1.030$ & $1.5070$ & $3.116$ & $\cdots$ & 9 \\
         \ion{Sc}{II} & $6245.621$ & $-1.022$ & $1.5070$ & $3.011$ & $\cdots$ & 15 \\
         \ion{Sc}{II} & $6320.832$ & $-1.816$ & $1.5004$ & $3.059$ & $\cdots$ & 7 \\
         
         \ion{V}{I} & $5670.831$ & $-0.430$ & $1.0806$ & $4.002$ & $\cdots$ & 21 \\
         \ion{V}{I} & $5737.027$ & $-0.740$ & $1.0636$ & $3.906$ & $\cdots$ & 16 \\
         \ion{V}{I} & $6081.461$ & $-0.610$ & $1.0509$ & $3.901$ & $\cdots$ & 10 \\
         \ion{V}{I} & $6224.520$ & $-2.010$ & $0.2866$ & $4.074$ & $\cdots$ & 21 \\
         \ion{V}{I} & $6251.802$ & $-1.370$ & $0.2866$ & $3.895$ & $\cdots$ & 21 \\
         \ion{V}{I} & $6274.659$ & $-1.700$ & $0.2670$ & $4.005$ & $\cdots$ & 6 \\
         \ion{V}{I} & $6285.162$ & $-1.540$ & $0.2753$ & $3.971$ & $\cdots$ & 12 \\
         
         \ion{Mn}{I} & $4502.204$ & $-0.345$ & $2.9197$ & $5.347$ & $+0.026$ & 15 \\
         \ion{Mn}{I} & $4671.647$ & $-1.675$ & $2.8884$ & $5.508$ & $+0.030$ & 15 \\
         \ion{Mn}{I} & $4739.072$ & $-0.607$ & $2.9408$ & $5.402$ & $+0.032$ & 10 \\
         \ion{Mn}{I} & $5377.626$ & $-0.166$ & $3.8437$ & $5.344$ & $+0.042$ & 12 \\
         \ion{Mn}{I} & $5399.500$ & $-0.345$ & $3.8530$ & $5.481$ & $+0.039$ & 10 \\
         \ion{Mn}{I} & $5413.697$ & $-0.647$ & $3.8590$ & $5.480$ & $+0.036$ & 6 \\
         
         \ion{Co}{I} & $4594.669$ & $-0.042$ & $3.6320$ & $4.604$ & $\cdots$ & 21 \\
         \ion{Co}{I} & $4792.854$ & $0.001$ & $3.2524$ & $4.800$ & $+0.111$ & 18 \\
         \ion{Co}{I} & $4813.471$ & $0.120$ & $3.2158$ & $4.849$ & $+0.116$ & 21 \\
         \ion{Co}{I} & $5301.017$ & $-2.000$ & $1.7104$ & $4.953$ & $+0.099$ & 16 \\
         \ion{Co}{I} & $5342.706$ & $0.741$ & $4.0208$ & $4.769$ & $\cdots$ & 21 \\
         \ion{Co}{I} & $5352.020$ & $0.060$ & $3.5764$ & $4.790$ & $\cdots$ & 20 \\
         \ion{Co}{I} & $5359.188$ & $0.244$ & $4.1494$ & $4.736$ & $\cdots$ & 21 \\
         \ion{Co}{I} & $5647.207$ & $-1.560$ & $2.2800$ & $4.883$ & $+0.102$ & 12 \\
         \ion{Co}{I} & $6814.958$ & $-1.900$ & $1.9557$ & $4.774$ & $+0.087$ & 10 \\
         
    \noalign{\smallskip}
    \hline
    \end{tabular}

    \footnotesize{Stellar parameters for the Sun are $T_\text{eff}=5777\pm18$\,K, $\log g=4.41\pm0.05$, and $\text{[Fe/H]}=0.0$\, dex ($\text{A(Fe)}=7.48\pm0.01$\, dex).\\
    References for the NLTE corrections: O -- \citet{Bergemann2021MNRAS.508.2236B}; Mn -- \citet{Bergemann2019A&A...631A..80B}; Co -- \citet{Bergemann2010MNRAS.401.1334B} and \citet{Voronov2022ApJ...926..173V}.}
    \label{tab:lines}
\end{table}

\begin{figure*}
	\includegraphics[width=2\columnwidth]{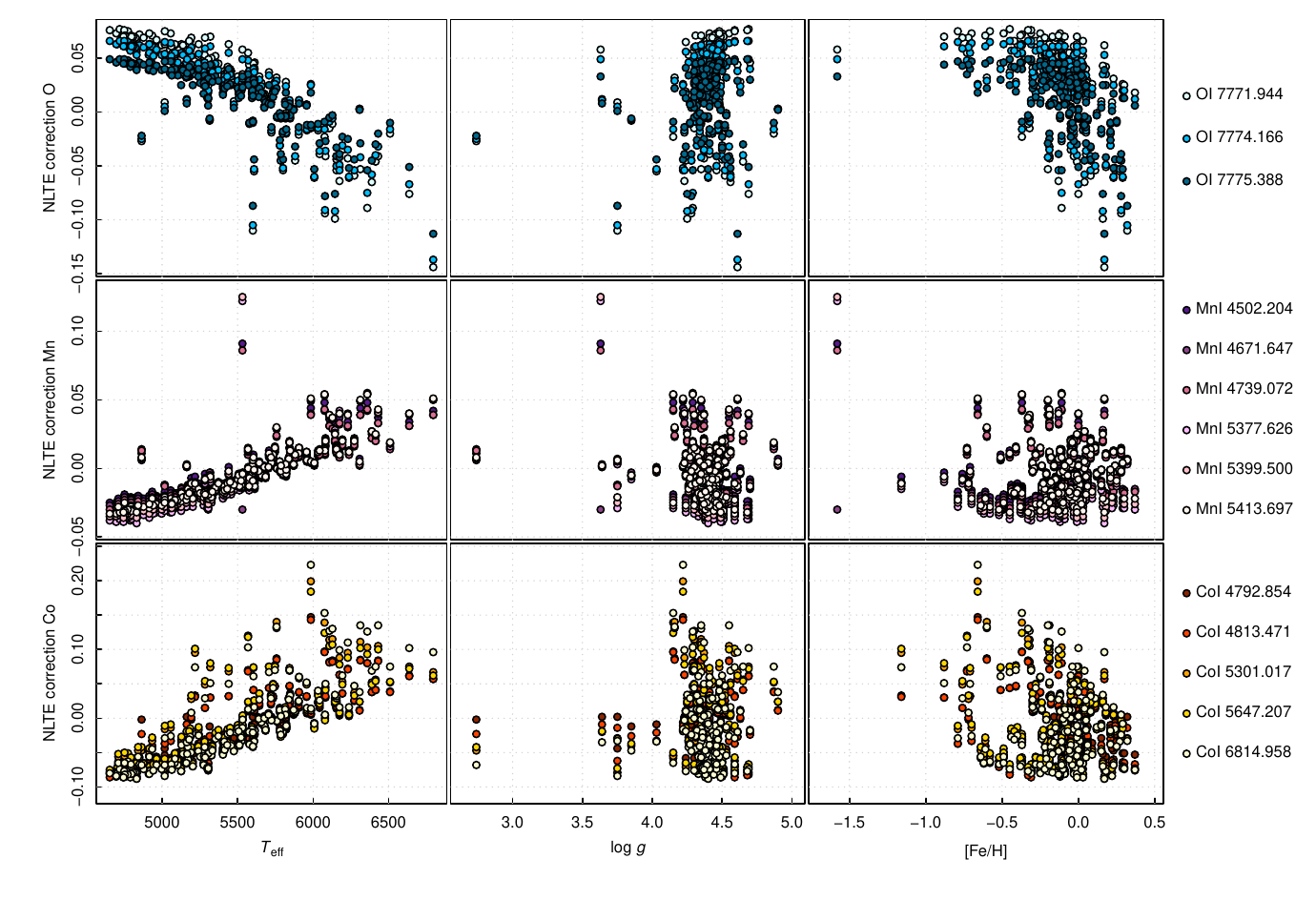}
    \caption{NLTE corrections for the O, Mn and Co lines versus de stellar parameters ($T_\text{eff}$, $\log g$, and [Fe/H]).}
    \label{fig:NLTE_effects}
\end{figure*}

In order to calculate the required abundances, we assumed the stellar parameters derived with \textsc{StePar} in \citetalias{Montes2018MNRAS.479.1332M} and used the \textit{EW} method, a grid of MARCS stellar atmospheric models \citep{Gustafsson2008A&A...486..951G}, and the 2019 version of the code MOOG\footnote{\url{https://www.as.utexas.edu/~chris/moog.html}} \citep{Sneden1973PhDT.......180S}. The \textit{EW} were computed using ARES\footnote{\url{https://github.com/sousasag/ARES}} \citep{Sousa2015A&A...577A..67S}. The final abundances were computed in a differential manner, in a line-by-line basis, with respect to our solar spectrum (using the asteroid Vesta) as in \citetalias{Montes2018MNRAS.479.1332M}, which minimise systematic uncertainties that may arise due to the analysis method and atomic data used. Out of 179 stars with available stellar parameters, we measured C abundances for 161 stars, O abundances for 173 stars, Sc abundances for all 179 stars, V abundances for 176 stars, Mn abundances for all 179 stars, and Co abundances for 178 stars. Since the \ion{C}{I} line at $5380.34$\,{\AA} presents an unkown blend, it was not considered for the final carbon abundance for stars with $T_\text{eff}<5200$\,K. The solar abundances for each line are listed in Table~\ref{tab:lines}.

\begin{figure}
    \centering
    \includegraphics[width=0.935\columnwidth]{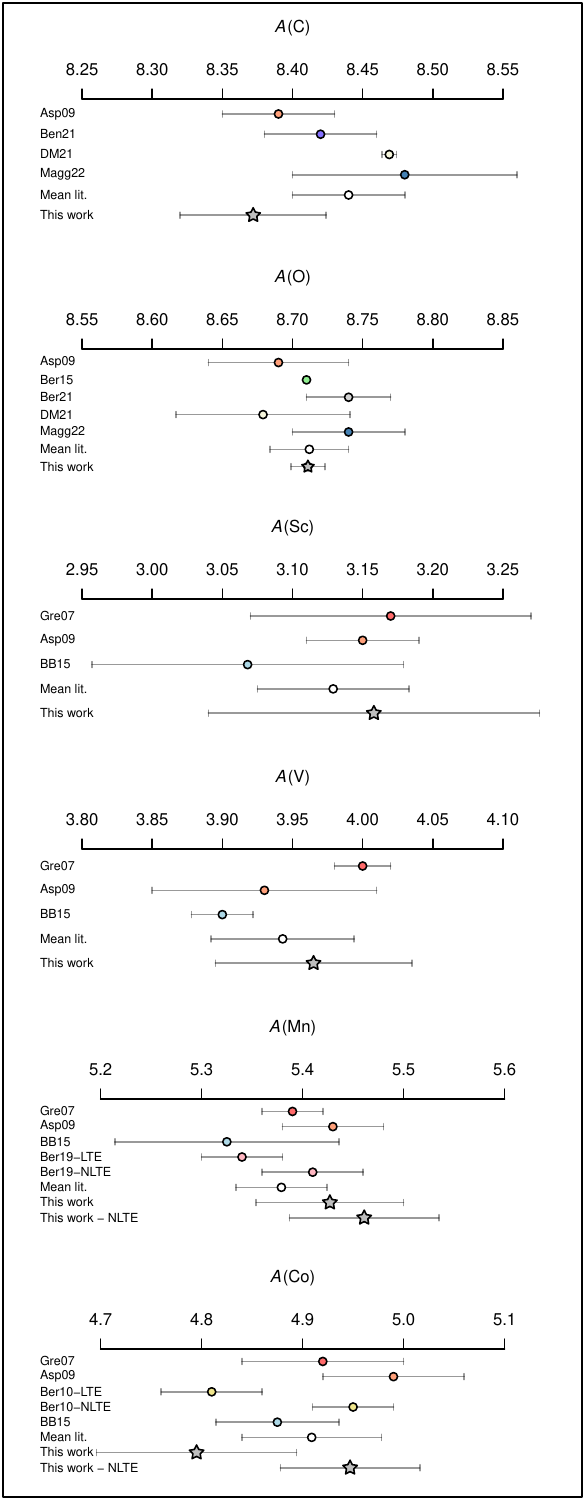}
    \caption{Comparison for our solar abundances (gray stars) with the literature: \citet{Grevesse2007SSRv..130..105G} (Gre07 -- red), \citet{Asplund2009ARA&A..47..481A} (Asp09 -- orange), \citet{Bergemann2010MNRAS.401.1334B} (Ber10 -- yellow), \citet{BertranDeLis2015A&A...576A..89B} (Ber15 -- green), \citet{Battistini2015A&A...577A...9B} (BB15 -- blue), \citet{Bergemann2019A&A...631A..80B} (Ber19 -- pink), \citet{Bensby2021A&A...655A.117B} (Ben21 -- purple), \citet{Bergemann2021MNRAS.508.2236B} (Ber21 -- light gray), \citet{Delgado-Mena2021A&A...655A..99D} (DM21 -- beige), \citet{Magg2022A&A...661A.140M} (Magg22 -- dark blue), and the mean value from the literature (Mean lit. -- white).}
    \label{fig:solar_abundances_literature}
\end{figure}

The odd-Z iron-peak elements Sc, V, Mn, and Co suffer from hyperfine structure, which splits their absorption lines into multiple components \citep{Jofre2017A&A...601A..38J,Heiter2021A&A...645A.106H}. \cite{Shan2021A&A...654A.118S} showed that the impact of hyperfine structure on certain absorption lines in cool photospheres can be more pronounced compared to Sun-like stars.
Therefore, it must be considered in order to derive precise abundances with the \textit{EW} method. In our analysis, we used the `abfind' driver within MOOG to derive carbon and oxygen abundances. However, to account for the HFS components of Sc, V, Mn, and Co, we used the `blends' driver, as in several previous studies \citep[see e.g.][]{daSilva2015A&A...580A..24D,Maldonado2015A&A...579A..20M,Maldonado2020A&A...644A..68M, Biazzo2015A&A...583A.135B, Biazzo2022A&A...664A.161B}.

Furthermore, some of the O, Mn and Co lines used in the abundance determination are affected by NLTE.
Since the MARCS models are based on a standard LTE approximation, the O abundances were later corrected for NLTE effects with the corrections given by \cite{Bergemann2021MNRAS.508.2236B}, the Mn abundances with corrections by \cite{Bergemann2019A&A...631A..80B}, and the Co abundances with corrections by \cite{Bergemann2010MNRAS.401.1334B} and collisional data from \cite{Voronov2022ApJ...926..173V}, using the web interface at \url{http://nlte.mpia.de/}. These corrections are based on the plane-parallel 1D \texttt{MAFAGS-OS} model atmosphere code \citep{Grupp2004A&A...420..289G}. We display the NLTE corrections against the stellar parameters in Fig.~\ref{fig:NLTE_effects}. We found a significant trend with $T_\text{eff}$, which is positive for Mn and Co NLTE corrections but negative for O NLTE corrections. According to \cite{Zhao2016ApJ...833..225Z}, Sc only presents large NLTE corrections in the low-metallicity regime ($\text{[Fe/H]}<-1.5$). We also compile the number of HFS components and the NLTE corrections in Table~\ref{tab:lines}. The atomic data of the HFS components are compiled in Tables~\ref{tab:HFS_Sc}--\ref{tab:HFS_Co}.

\begin{table*}
    \centering
    \caption{Average of the different contributions to the total abundance uncertainties, analysing the abundance sensitivies to changes of each stellar parameter.}
    \begin{tabular}{llcccccc} \hline
    \noalign{\smallskip}
         Abundance & Stars & Line-to-line scatter & $\Delta T_\text{eff}$ & $\Delta \text{[Fe/H]}$ & $\Delta\log g$ & $\Delta\xi$ & Total error \\
          & & (dex) & (dex) & (dex) & (dex) & (dex) & (dex) \\ 
    \noalign{\smallskip}
    \hline
    \noalign{\smallskip}

    [C/H] & F-type stars & $\pm0.034$ & $\pm0.018$ & $\pm0.001$ & $\mp 0.026$ & $\pm0.001$ & $\pm0.048$ \\
     & G-type stars & $\pm0.081$ & $\pm0.023$ & $\pm0.001$ & $\mp0.027$ & $\pm0.001$ & $\pm0.092$ \\
     & K-type stars & $\pm0.111$ & $\pm0.062$ & $\pm0.002$ & $\mp0.066$ & $\pm0.001$ & $\pm0.143$ \\

     [O/H] & F-type stars & $\pm0.060$ & $\pm0.024$ & $\pm0.001$ & $\mp0.015$ & $\pm0.005$ & $\pm0.073$ \\
      & G-type stars & $\pm0.073$ & $\pm0.034$ & $\pm0.001$ & $\mp0.016$ & $\pm0.003$ & $\pm0.098$ \\
      & K-type stars & $\pm0.094$ & $\pm0.099$ & $\pm0.001$ & $\mp0.056$ & $\pm0.004$ & $\pm0.144$ \\

      [Sc/H] & F-type stars & $\pm0.083$ & $\pm0.009$ & $\pm0.004$ & $\mp0.023$ & $\pm0.006$ & $\pm0.098$ \\
      & G-type stars & $\pm0.075$ & $\pm0.008$ & $\pm0.004$ & $\mp0.022$ & $\pm0.006$ & $\pm0.089$ \\
      & K-type stars & $\pm0.133$ & $\pm0.023$ & $\pm0.008$ & $\mp0.042$ & $\pm0.018$ & $\pm0.153$ \\

      [V/H] & F-type stars & $\pm0.163$ & $\pm0.031$ & $\pm0.001$ & $\mp0.003$ & $\pm0.001$ & $\pm0.173$ \\
      & G-type stars & $\pm0.065$ & $\pm0.036$ & $\pm0.000$ & $\mp0.000$ & $\pm0.002$ & $\pm0.079$ \\
      & K-type stars & $\pm0.057$ & $\pm0.098$ & $\pm0.001$ & $\mp0.010$ & $\pm0.027$ & $\pm0.126$\\

      [Mn/H] & F-type stars & $\pm0.113$ & $\pm0.020$ & $\pm0.001$ & $\mp0.001$ & $\pm0.002$ & $\pm0.116$ \\
      & G-type stars & $\pm0.089$ & $\pm0.020$ & $\pm0.001$ & $\mp0.001$ & $\pm0.006$ & $\pm0.095$ \\
      & K-type stars & $\pm0.097$ & $\pm0.034$ & $\pm0.006$ & $\mp0.006$ & $\pm0.018$ & $\pm0.107$ \\

      [Co/H] & F-type stars & $\pm0.066$ & $\pm0.023$ & $\pm0.001$ & $\mp0.000$ & $\pm0.001$ & $\pm0.077$ \\
      & G-type stars & $\pm0.056$ & $\pm0.016$ & $\pm0.002$ & $\mp0.009$ & $\pm0.003$ & $\pm0.061$ \\
      & K-type stars & $\pm0.074$ & $\pm0.012$ & $\pm0.008$ & $\mp0.038$ & $\pm0.011$ & $\pm0.087$ \\

    \noalign{\smallskip}
    \hline
    \end{tabular}
    \label{tab:error_contributions}
\end{table*}

\subsection{Abundance uncertainties}

In \citetalias{Montes2018MNRAS.479.1332M}, the abundance uncertainties were computed as the scatter of the abundances obtained from each line for the same atomic species. In this work, we proceeded as in \citet{Battistini2015A&A...577A...9B} (hereafter \citetalias{Battistini2015A&A...577A...9B}), \citet{Delgado-Mena2021A&A...655A..99D} (hereafter \citetalias{Delgado-Mena2021A&A...655A..99D}), and others, taking the uncertainties in the stellar parameters ($T_\text{eff}$, $\log g$, $\xi$, and [Fe/H]) into consideration.
The abundance uncertainty due to the errors on stellar parameters were estimated by calculating the abundance differences when one of each of the stellar parameters was modified by its individual error given in Table B.3 of \citetalias{Montes2018MNRAS.479.1332M}. The final abundance errors are given by the quadratic sum of these individual errors and the line-to-line scatter. The average of these contributions to the total uncertainty for each abundance and spectral type is given in Table~\ref{tab:error_contributions}. As the analysis is differential relative to the Sun in a line-by-line basis, systematic errors should be minimised.

Oxygen and carbon can form several molecules, such as the CO, in the atmospheres of late-type stars. Therefore, abundances of O and C are bound via chemical equilibrium. As demonstrated by \cite{Pavlenko2019A&A...621A.112P}, the interdependence of these abundances in solar-type stars is negligible, within the errorbars.

\subsection{Solar abundances}

Since our analysis is differential relative to the Sun in a line-by-line basis, we also derived the abundances for the Sun, obtaining $A(\text{C})_\odot = 8.372\pm0.052$\,dex, $A(\text{O})_\odot = 8.711\pm0.012$\,dex, $A(\text{Sc})_\odot = 3.158\pm0.118$\,dex, $A(\text{V})_\odot = 3.965\pm0.070$\,dex, $A(\text{Mn})_\odot = 5.427\pm0.073$\,dex, and $A(\text{Co})_\odot = 4.795\pm0.099$\,dex. The comparison between our solar abundances and the ones from the literature is shown in Fig. \ref{fig:solar_abundances_literature}. The differences may be due to the use of different line lists, and the assumptions made, such us LTE, 1D models, etc.

\section{Results and discussion}
\label{Sect_results}

\subsection{Carbon and oxygen abundances}

\begin{figure*}
    \begin{tabular}{cc}
       \includegraphics[width=0.99\columnwidth]{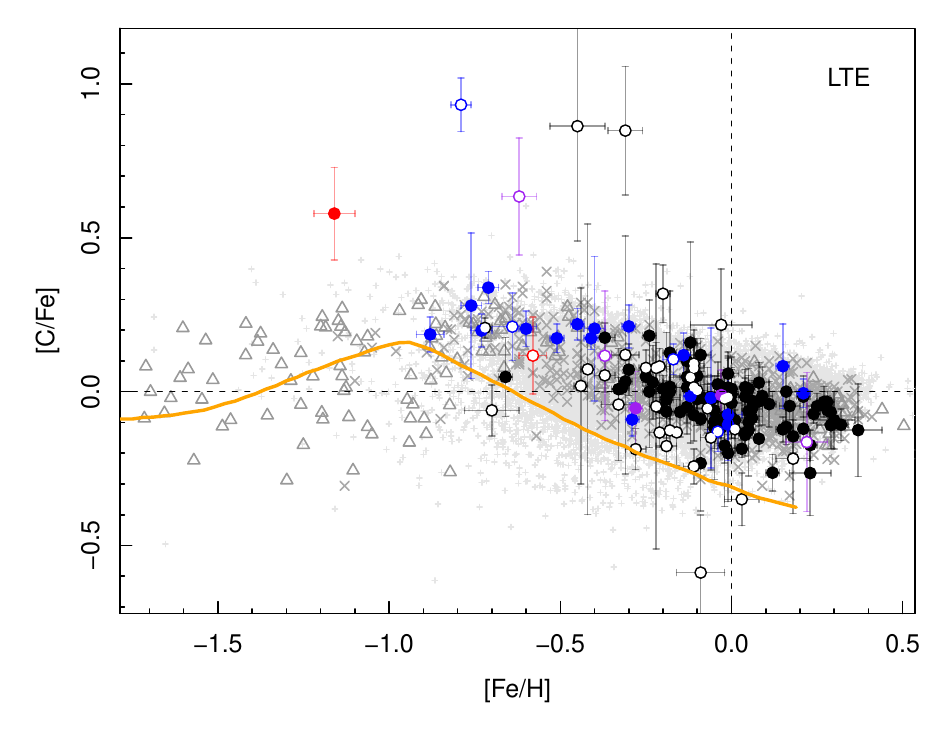} & 
       \includegraphics[width=0.99\columnwidth]{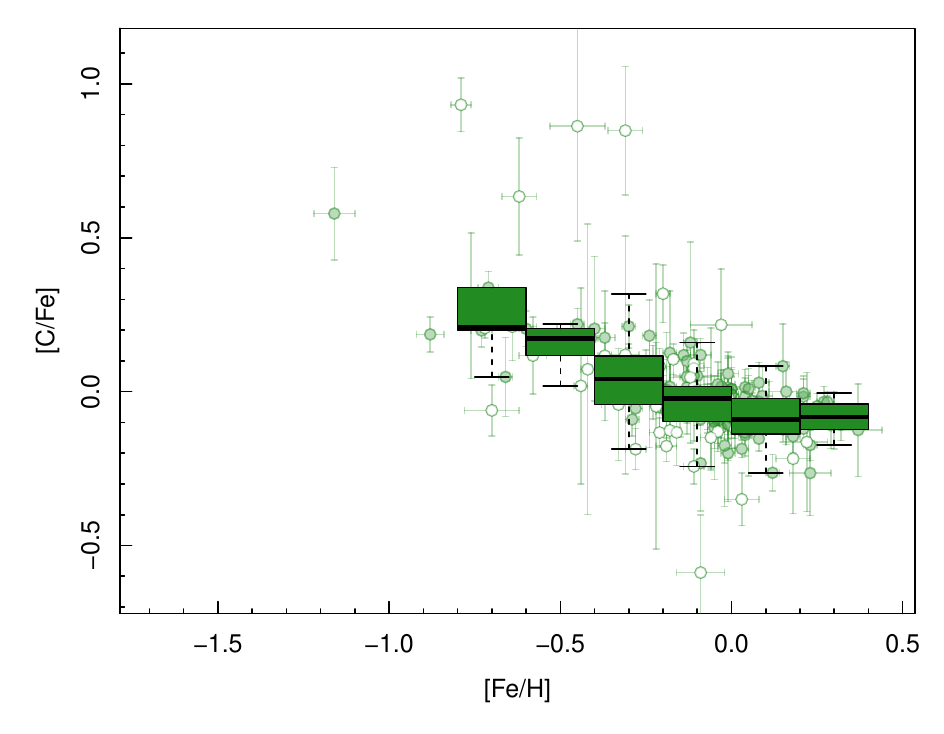} \\

       \includegraphics[width=0.99\columnwidth]{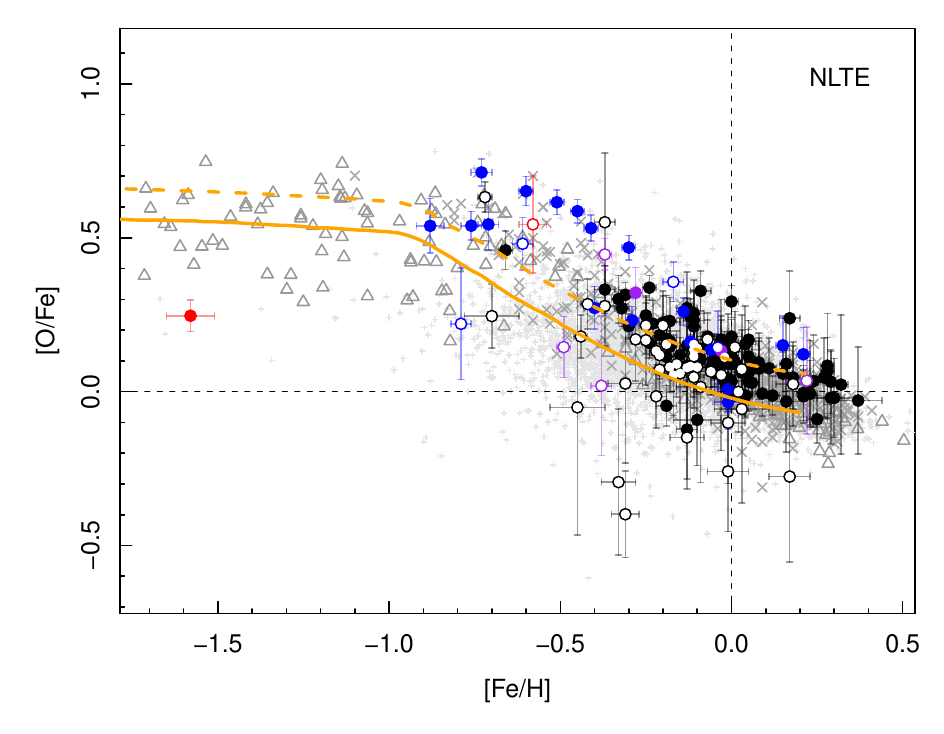} &
       \includegraphics[width=0.99\columnwidth]{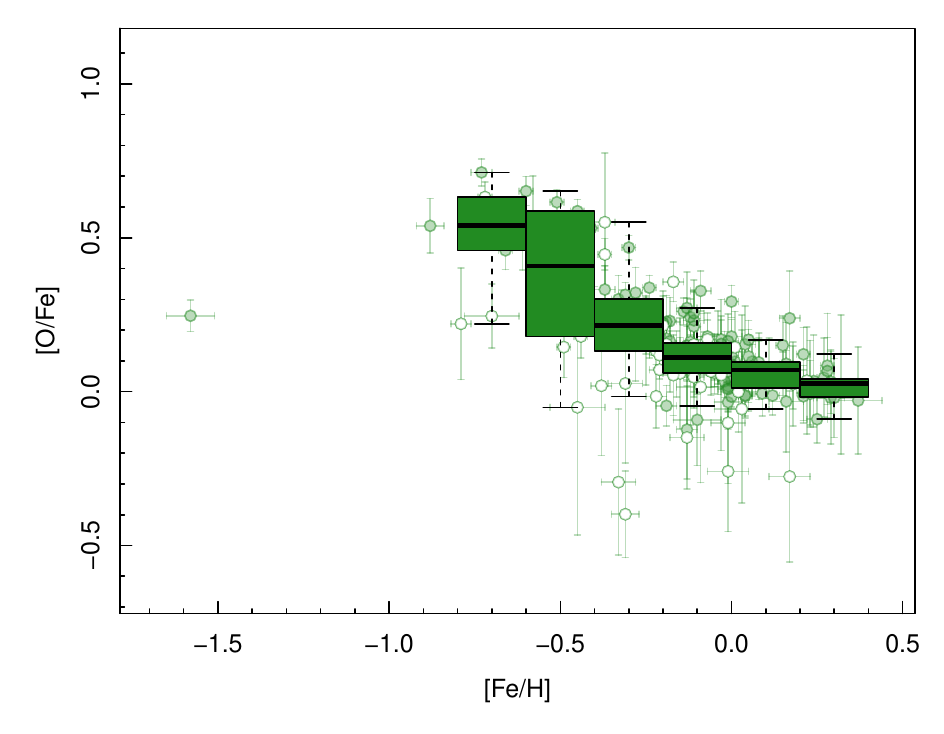} \\

       \includegraphics[width=0.99\columnwidth]{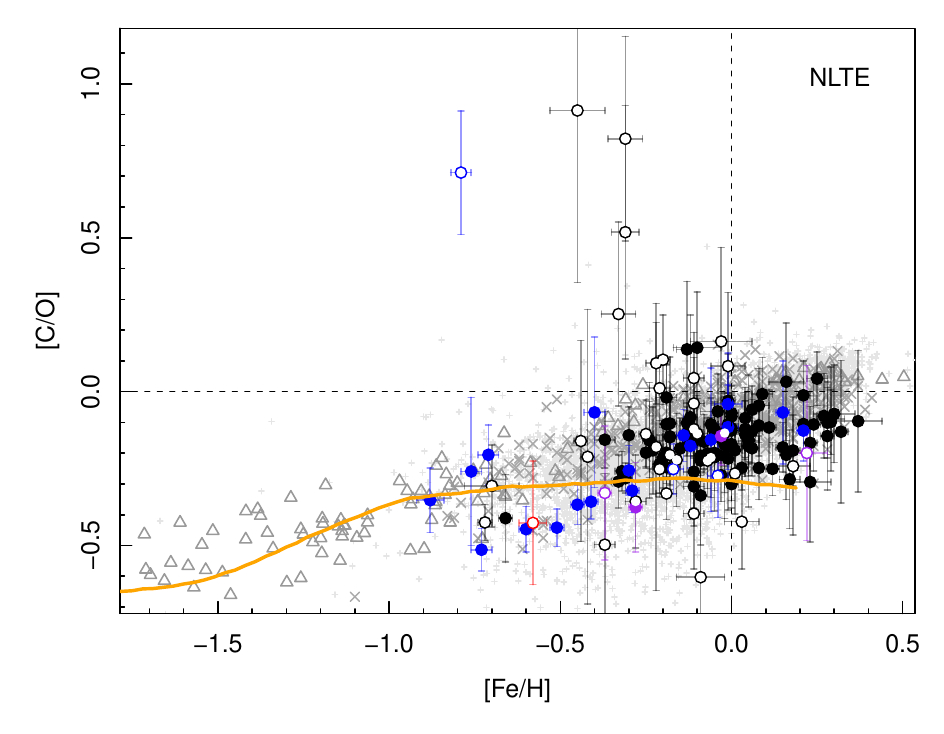} &
       \includegraphics[width=0.99\columnwidth]{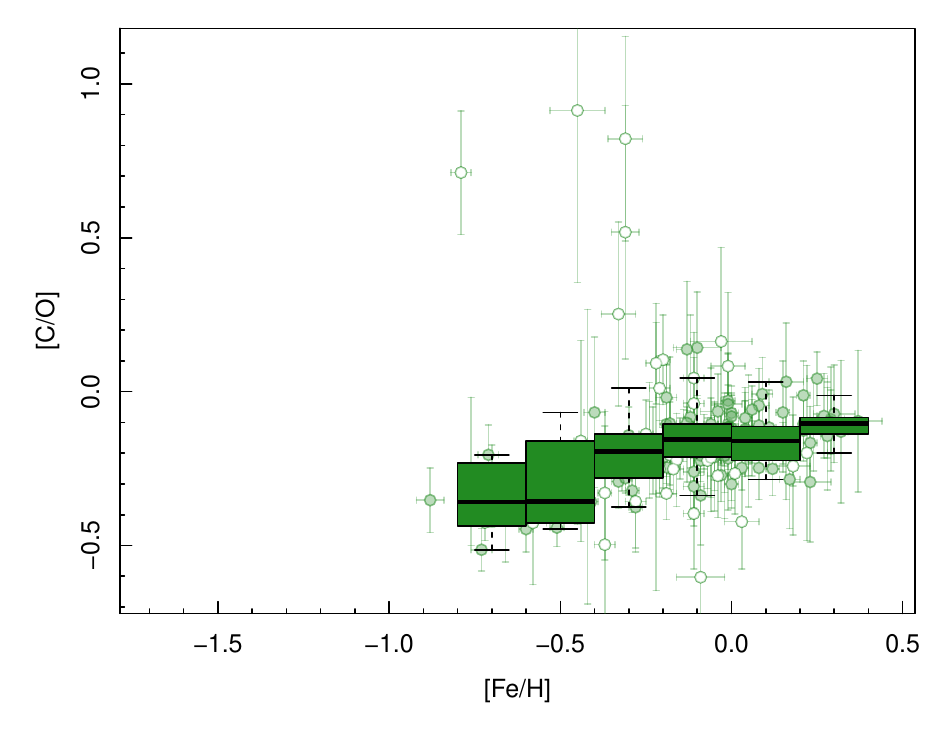} \\
    \end{tabular}
    \includegraphics[width=1.7\columnwidth]{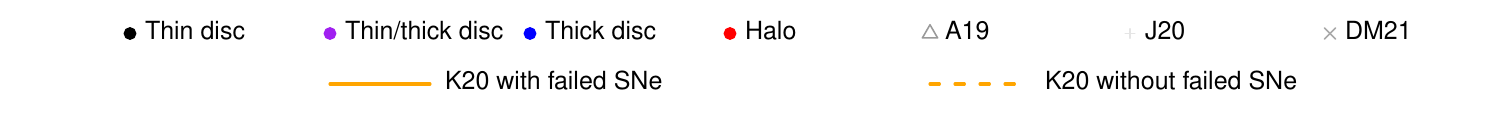}

    \caption{Left panels: Abundance ratios of [C/Fe], [O/Fe], and [C/O] versus [Fe/H], in comparison with \citet{Amarsi2019A&A...630A.104A} (A19) using the 1D/NLTE results, \citet{Jonsson2020AJ....160..120J} (J20), and \citetalias{Delgado-Mena2021A&A...655A..99D}. Colours for different populations: black -- thin disc, blue -- thick disc, purple -- thin/thick disc, red -- halo. Open circles: stars with $T_\text{eff}<5200$\,K; filled circles: stars with $T_\text{eff}\geq 5200$\,K. Models of GCE from \citet{Kobayashi2020ApJ...900..179K} with failed SNe at $>30M_\odot$ (solid orange line) and without failed SNe (dotted orange line) are overplotted.
    Right panels: Box and whisker plots of [C/Fe], [O/Fe], and [C/O] versus [Fe/H].}
    \label{fig:CFe_OFe}
\end{figure*}

We display the distributions of [C/Fe], [O/Fe], and [C/O] as a function of metallicity in left panels of Fig.~\ref{fig:CFe_OFe}. Previous papers (e.g. \citealt{BertranDeLis2015A&A...576A..89B}, \citetalias{Delgado-Mena2021A&A...655A..99D}) have set the limit for reliable abundances of C and O at 5200\,K; for stars below this effective temperature, abundances provided by different lines start to disagree. In Fig.~\ref{fig:CFe_OFe} we distinguish stars with $T_\text{eff}<5200$\,K as open circles, and note that most of the outliers belong to this category. Our analysis revealed increasing [C/Fe] and [O/Fe] ratios with decreasing [Fe/H], while the [C/O] ratio increase towards higher metallicities. Moreover, adopting the Galactic populations derived in \citetalias{Montes2018MNRAS.479.1332M} from the Galactocentric space velocities (see \citealt{Montes2001MNRAS.328...45M, Bensby2003A&A...410..527B, Bensby2005A&A...433..185B}), namely halo, thick disc, thick-to-thin transition disc, and thin disc, we observed that thick disc stars present higher [C/Fe] and [O/Fe] values than thin disc stars in the common metallicity interval. Furthermore, we overplot (with orange lines) the GCE models from \citet{Kobayashi2020ApJ...900..179K} (hereafter K20 model), which assumed failed supernovae (SNe) for stars with initial masses higher than $30M_\odot$. For the [O/Fe] abundance we also included a K20 model without failed SNe. To better examine the trends of these ratios with increasing metallicity, we show box and whisker plots in right panels of Fig. \ref{fig:CFe_OFe}.

Our findings on the [C/Fe] trend are consistent with previous studies by \cite{Delgado-Mena2010ApJ...725.2349D,Delgado-Mena2021A&A...655A..99D} and \cite{Amarsi2019A&A...630A.104A}. Other authors reported similar results \citep[e.g.][]{Takeda2005PASJ...57...65T,Reddy2006MNRAS.367.1329R,Gonzalez-Hernandez2010ApJ...720.1592G,Gonzalez-Hernandez2013A&A...552A...6G,Nissen2014A&A...568A..25N,Buder2019A&A...624A..19B,Franchini2020ApJ...888...55F,Stonkute2020AJ....159...90S}, showing increasing [C/Fe] ratios with decreasing metallicity down to $\text{[Fe/H]}\sim -0.8$\,dex, after which the trend flattens and then decreases for halo stars. Notably, the halo star HD 149414 Aa,Ab exhibits a higher carbon abundance than expected based on the general trend. This overabundance could be attributed to the fact that this star is a single-lined spectroscopic binary (SB1; \citealt{Latham2002AJ....124.1144L}) with $T_\text{eff}=5217\pm57$\,K, in close proximity to the 5200\,K limit.

Several studies have examined the distribution of [O/Fe] over [Fe/H], and found that the [O/Fe] ratio rises with decreasing metallicity, albeit with varying slopes, and reaches a plateau around $\text{[Fe/H]}=-0.8$\,dex. While some of them found just a shallow decline in the [O/Fe] trend at high metallicities \citep[e.g.][]{Bensby2004A&A...415..155B,Bensby2014A&A...562A..71B,Takeda2005PASJ...57...65T,Ramirez2007A&A...465..271R,Petigura2011ApJ...735...41P,Franchini2021AJ....161....9F}, other studies suggested a flattening at solar or super-solar metallicities (e.g. \citet{Bensby2004A&A...415..155B,Ramirez2013ApJ...764...78R,BertranDeLis2015A&A...576A..89B}, \citetalias{Delgado-Mena2021A&A...655A..99D}). These differences could be attributed to the use of different oxygen indicators or whether the author included or neglected NLTE effects, inelastic collisions with hydrogen, etc. In this case, the red giant branch star BD+80 245 ($\text{[Fe/H]}= -1.58\pm0.07$\,dex) shows a low O abundance when compared to the Galactic trend. The low metallicity, which implies weak lines, and low signal-to-noise ratio of this star ($\text{S/N}\sim 60$) could explain the underabundance obtained with the EW method. Moreover, we note that the thick-disc stars present higher C and O abundances than the thin-disc stars at equivalent [Fe/H] values.


These results for the [C/Fe] and [O/Fe] trends discussed above are consistent with what it is expected from Galactic chemical evolution. Both [C/Fe] and [O/Fe] decrease with increasing [Fe/H], which reflects the increasing relevance of SNe Ia with time, fueling the increase in iron, relative to SNe II. However, the slopes of both trends are different, with the carbon following iron more closely than oxygen, due to their different production sites. Oxygen is produced by SNe II from massive progenitors \citep{Woosley1995ApJS..101..181W}. Carbon, on the other hand, is produced by synthesis in stellar interiors, then dredged up from cores and released into the interstellar medium by massive stellar winds and radiation pressure \citep{Gustafsson1999A&A...342..426G}. However, the lower-mass post-AGB stars contribute as well, and their relative importance is still a matter of debate \citep[see][and references therein]{Franchini2020ApJ...888...55F}. Therefore, the enrichment of carbon is delayed in time with respect to oxygen, explaining the different slopes.
As can be seen in Fig.~\ref{fig:CFe_OFe}, our results follow the general trend of the K20 models but, as stated by \citet{Kobayashi2020ApJ...900..179K}, the predicted [C/Fe] is $\sim 0.1$--$0.2$\,dex lower than the observational data, with a steeper decrease for $\text{[Fe/H]}>-1$. However, as shown by \citet{Delgado-Mena2021A&A...655A..99D} in their Fig.~3, including the contribution of Wolf–Rayet stars to the carbon production can increase the [C/Fe] abundance (Kobayashi et al. in prep.). In the case of [O/Fe], the model follows the general trend of the observations, but with lower abundances for $\text{[Fe/H]}>-1$\,dex. In this case, we also added a K20 model without failed SNe (dashed orange line), which seems to better reproduce our results for $\text{[Fe/H]}>-1$\,dex.

\begin{figure}
    \centering
    \includegraphics[width=\columnwidth]{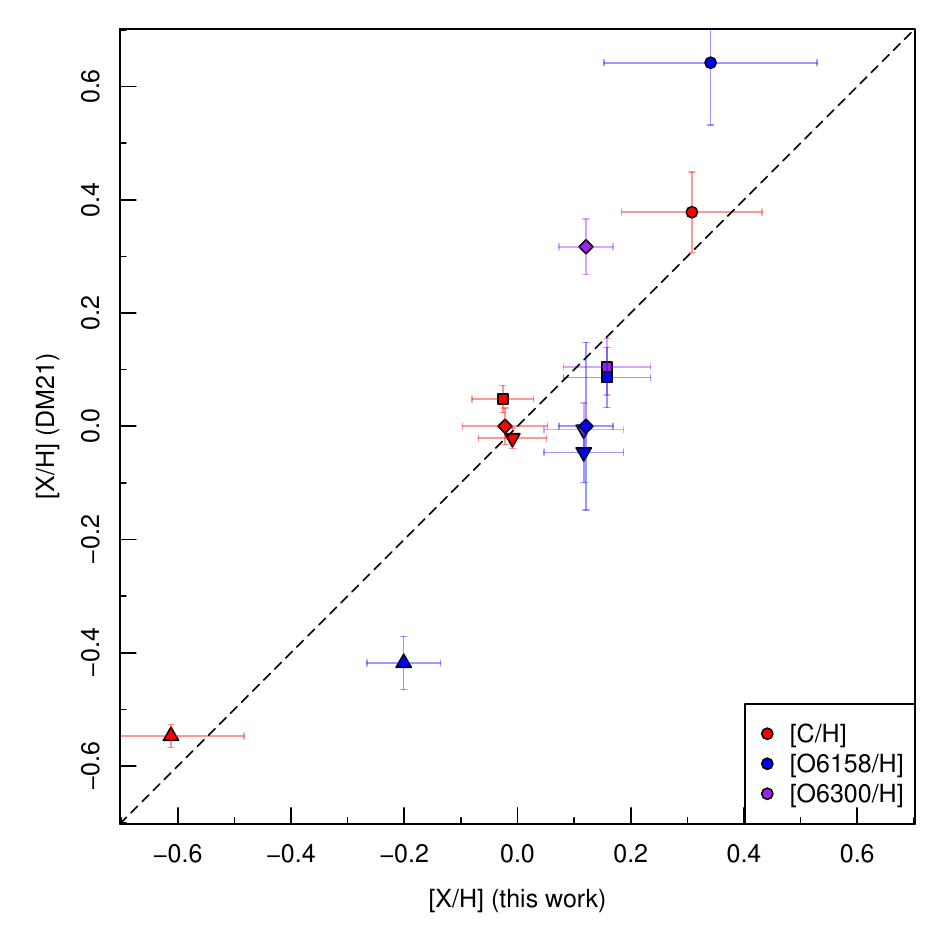}

    \includegraphics[width=\columnwidth]{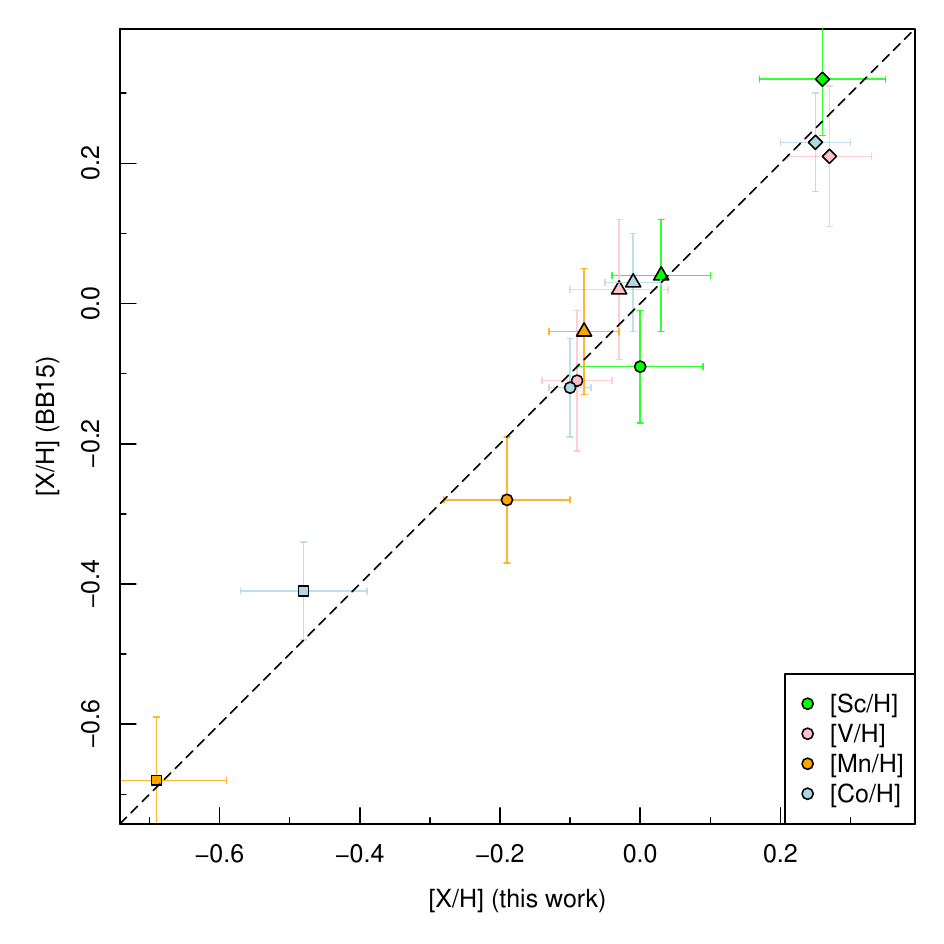}
    
    \caption{Upper panel: Comparison between the C and O abundances obtained by \citetalias{Delgado-Mena2021A&A...655A..99D} (who reported the \ion{O}{I} 6158 and 6300\,{\AA} abundances independently), and by this work for the stars in common: HD~8389 A (circle), HD~11964 A (square), HD~40397 A (diamond), HD~59984 (point-up triangle), and HD~222582 A (point-down triangle).
    Lower panel: Comparison between the Sc, V, Mn, and Co abundances by \citetalias{Battistini2015A&A...577A...9B} and by this work, only in LTE, for the stars in common: HD~40397 A (circle), HD~114606 A (square), HD~190360 A (diamond), and HD~222582 A (point-up triangle).}
    \label{fig:comparison_BB15_DM21}
\end{figure}

In the upper panel of Fig. \ref{fig:comparison_BB15_DM21} we compare the [C/H] and [O/H] ratios for the five stars in common with \citetalias{Delgado-Mena2021A&A...655A..99D}. \citetalias{Delgado-Mena2021A&A...655A..99D} provide the abundances of the \ion{O}{I} 6158 and 6300\,{\AA} lines independently, hence we represent our O abundances (using the \ion{O}{I} infrared triplet) against both results. The abundances of these stars closely correspond to those provided by \citetalias{Delgado-Mena2021A&A...655A..99D}, especially in the case of carbon abundances, which exhibit a low scatter.

\subsection{Odd-Z iron-peak abundances}

\begin{figure*}
    $\begin{tabular}{cc}
        \includegraphics[width=0.99\columnwidth]{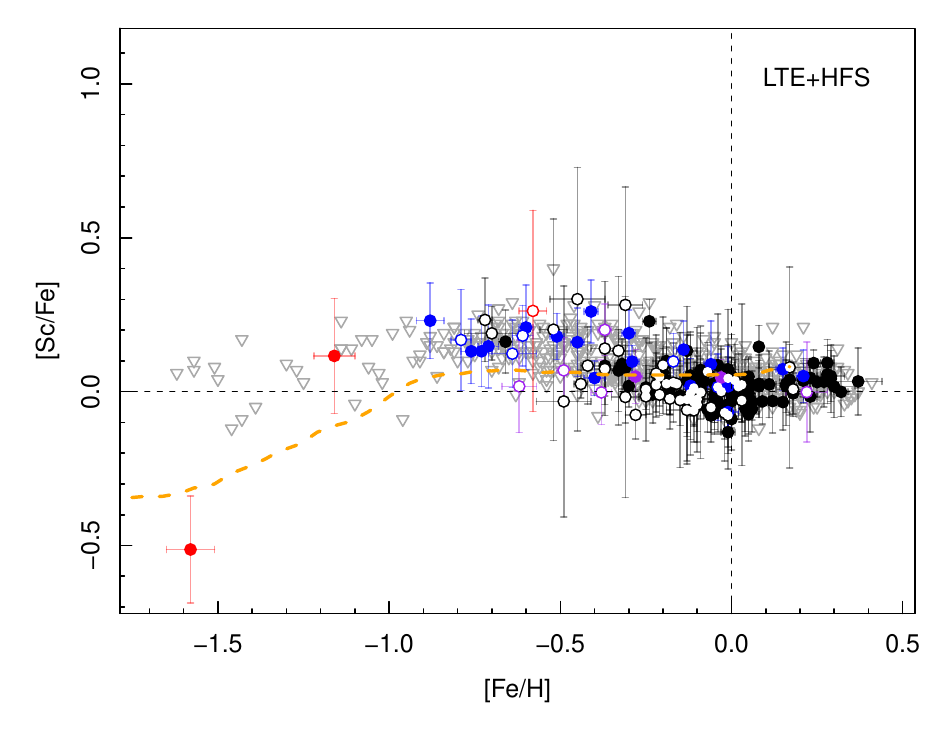} & 
        \includegraphics[width=0.99\columnwidth]{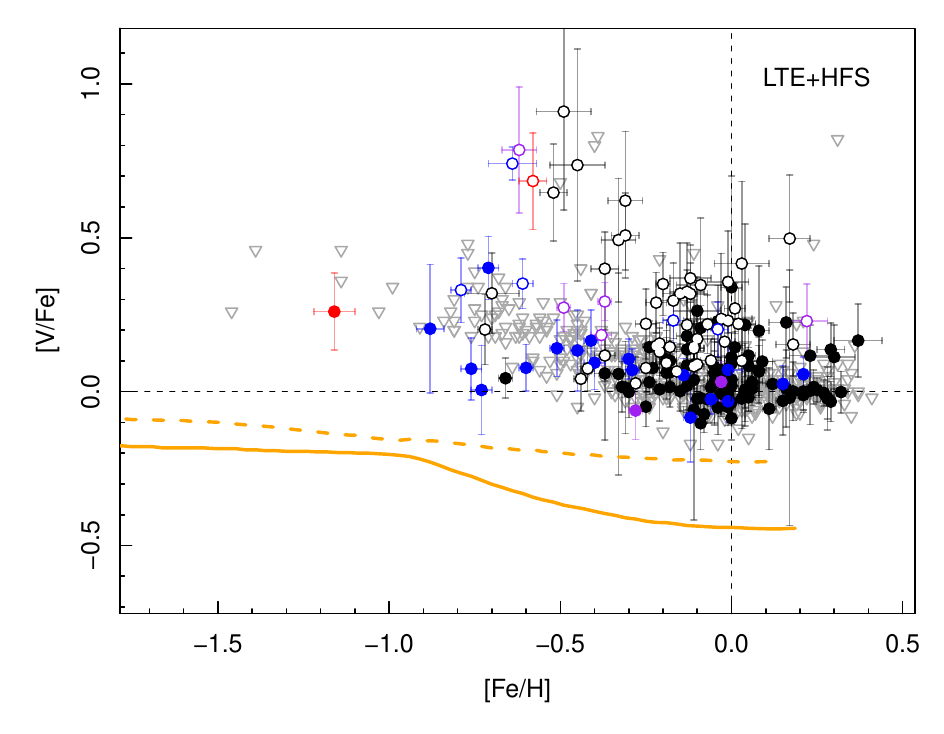} \\
        \includegraphics[width=0.99\columnwidth]{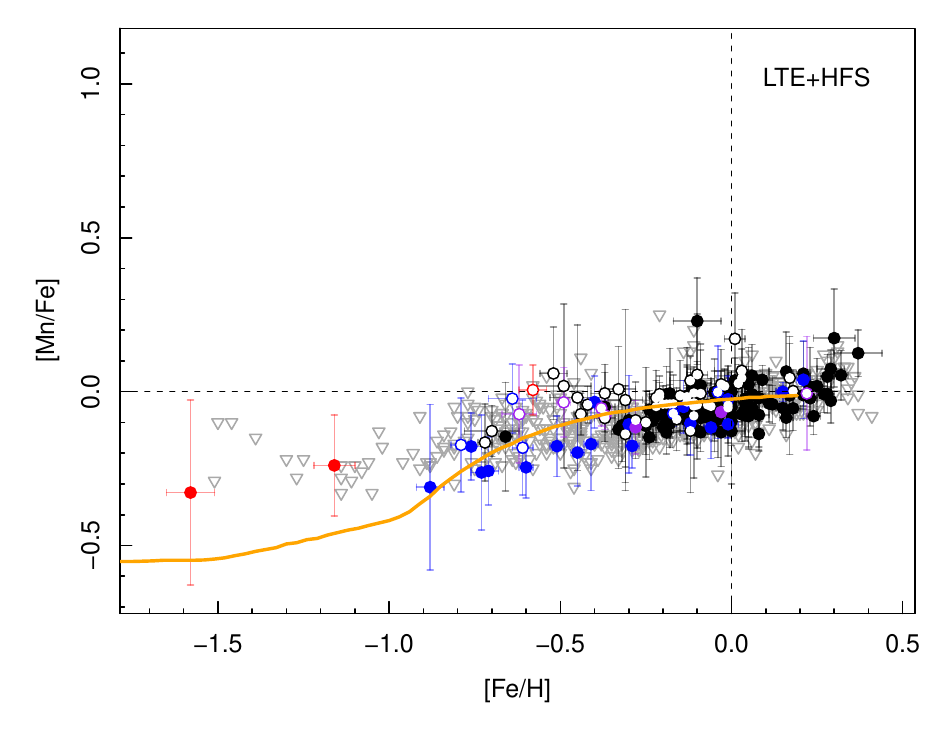} & 
        \includegraphics[width=0.99\columnwidth]{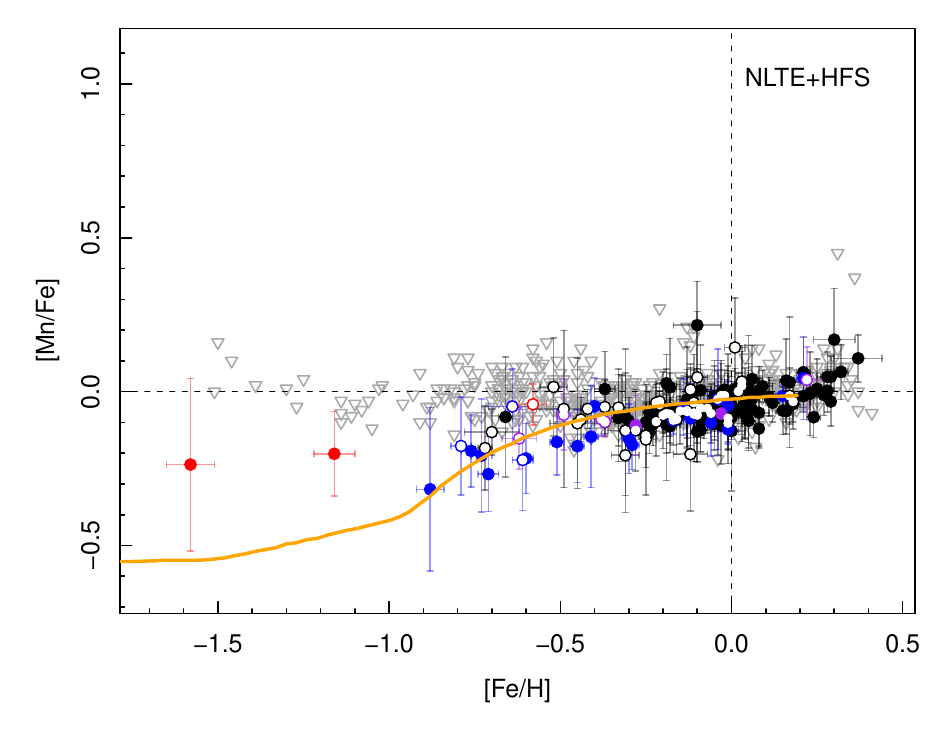} \\
        \includegraphics[width=0.99\columnwidth]{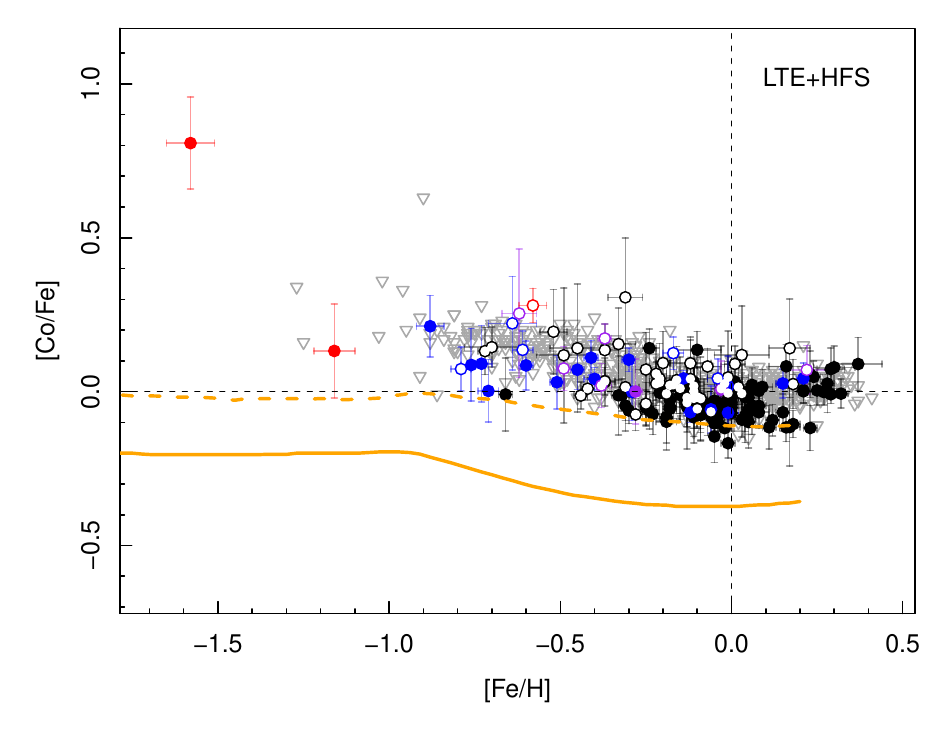} & 
        \includegraphics[width=0.99\columnwidth]{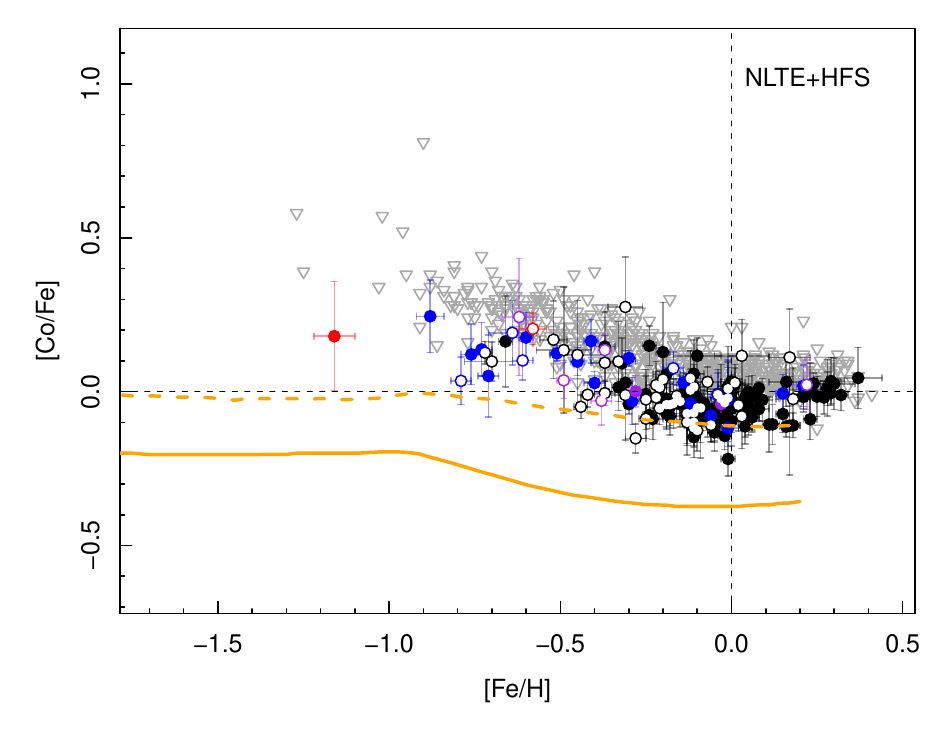} \\
    \end{tabular}$

    \includegraphics[width=1.7\columnwidth]{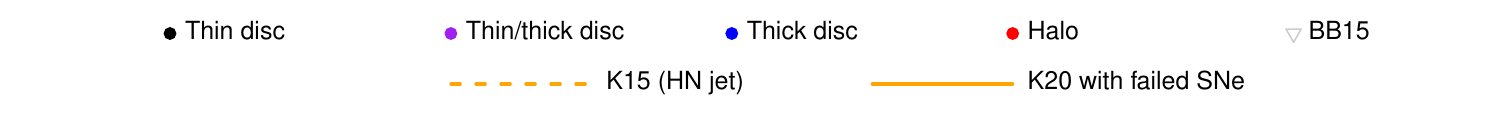}
    
    \caption{Abundance ratios of [Sc/Fe], [V/Fe], [Mn/Fe] (LTE and NLTE), and [Co/Fe] (LTE and NLTE) versus [Fe/H], in comparison with \citetalias{Battistini2015A&A...577A...9B}. Colours for different populations: black -- thin disc, blue -- thick disc, cyan -- thin/thick disc, red -- halo. Open circles: stars with $T_\text{eff}<5200$\,K; filled circles: stars with $T_\text{eff}\geq 5200$\,K. Models of GCE from \citet{Kobayashi2011MNRAS.414.3231K}, modified to include a hypernova jet effect (dotted orange line), and from \citet{Kobayashi2020ApJ...900..179K} with failed SNe at $>30M_\odot$ (solid orange line) are overplotted.}
    \label{fig:hfsFe}
\end{figure*}

We show the distribution of [X/Fe] ratios for the odd-Z iron-peak elements versus the metallicity in Fig.~\ref{fig:hfsFe}, in comparison with the results of \citetalias{Battistini2015A&A...577A...9B}. We also overplot in orange a modification of the GCE models from \citet{Kobayashi2011MNRAS.414.3231K} that includes a hypernova jet effect (dashed line; hereafter K15, see \citealt{Sneden2016ApJ...817...53S,Zhao2016ApJ...833..225Z,Kobayashi2020ApJ...900..179K}), and the K20 models (solid line). The [Sc/Fe] vs. [Fe/H] trend behave like the $\alpha$-elements, which suggests a common production site, and it is in good agreement with the results by \citet{Nissen2000A&A...353..722N}, \citet{Brewer2006AJ....131..431B}, \citet{Neves2009A&A...497..563N}, \citet{Adibekyan2012A&A...545A..32A}, \citet{Tabernero2012A&A...547A..13T}, and \citetalias{Battistini2015A&A...577A...9B}. We observe higher abundances for the thick disc stars in contrast to the thin disc. 
The K20 model predicts $\text{[Sc/Fe]}\sim-1$\,dex for our metallicity range, not visible in our panel. The K15 model provides a better fit to the observations, though it does not show the weak decrease of [Sc/Fe] with increasing metallicity for $\text{[Fe/H]}>-1$\,dex. In theoretical models, both Sc and V are consistently found to be underproduced across all metallicity ranges compared to observational data. The abundances of both elements could potentially be influenced by the neutrino process, a factor not accounted for in these GCE models.


The [V/Fe] vs. [Fe/H] trend also emulates an $\alpha$-element, with solar abundance at $\text{[Fe/H]}=0$ and a rise towards lower metallicities, but with significantly more scatter than the other elements when including the stars with $T_\text{eff}<5200$\,K. Vanadium is an intricate element that presents biases and higher dispersion for the cooler objects. For this reason, some studies established a cutoff limit around 5300\,K (for example, see \citealt{Neves2009A&A...497..563N} and \citealt{Adibekyan2012A&A...545A..32A}). The general trend is consistent with \citet{Brewer2006AJ....131..431B}, \citet{Neves2009A&A...497..563N}, \citet{Adibekyan2012A&A...545A..32A}, \citet{Tabernero2012A&A...547A..13T}, and \citetalias{Battistini2015A&A...577A...9B}. 
As in the case of Sc, V is underproduced by the GCE models. The K20 model shows a plateau and a weak decrease toward higher metallicities for $\text{[Fe/H]}>-1$\,dex, a similar behaviour that the observations, despite the $0.4$--$0.5$\,dex bias.


The [Mn/Fe] vs. [Fe/H] trend, assuming LTE, seems to work opposite to an $\alpha$-element \citep{Gratton1989A&A...208..171G}, i.e. it decreases steadily towards lower metallicities and shows sub-solar abundances for all metallicities up to $\text{[Fe/H]}=0$. This trend can be explained by models in which Mn is produced in SNII at metallicities below $\text{[Fe/H]}<-1$ \citep{Tsujimoto1998ApJ...508L.151T}, while at $\text{[Fe/H]}>-1$ the increasing [Mn/Fe] ratio with increasing [Fe/H] is due to a contribution from SNIa \citep{Kobayashi2016Natur.540..205K}. This trend is in good agreement with the results by \citet{Nissen2000A&A...353..722N}, \citet{Brewer2006AJ....131..431B}, \citet{Feltzing2007A&A...467..665F}, \citet{Neves2009A&A...497..563N}, \citet{Adibekyan2012A&A...545A..32A}, \citet{Tabernero2012A&A...547A..13T}, and \citetalias{Battistini2015A&A...577A...9B}. Nonetheless, when taking into account NLTE corrections, our trend presents a different behaviour compared to the one by \citetalias{Battistini2015A&A...577A...9B}, whose trend becomes virtually flat over the metallicity range studied. As the authors discussed, no models have been able to explain this flattening.
The K20 model predicts a plateau around $\text{[Mn/Fe]}\sim0.55$\,dex for the low-metallicity regime, and then an increase above $\text{[Fe/H]}>-1$\,dex.


Co shows a trend similar to $\alpha$-elements, with over-abundant [Co/Fe] at low metallicities. The halo star BD+80 245 (G0\, IV) seems to be Co-rich. Assuming LTE, this trend is in agreement with \citet{Brewer2006AJ....131..431B}, \citet{Neves2009A&A...497..563N}, \citet{Adibekyan2012A&A...545A..32A}, \citet{Tabernero2012A&A...547A..13T}, and \citetalias{Battistini2015A&A...577A...9B}. However, we find some differences when considering NLTE corrections in comparison with \citetalias{Battistini2015A&A...577A...9B}, who reported higher Co abundances for a given metallicity.
The K20 model reproduces the $\alpha$-element trend, but provides values $\sim0.3$\,dex lower than the observations. However, when including the hypernova jet effects (K15 model), the prediction is closer to the data.

Since our results for Mn and Co abundances assuming LTE are in good agreement with those by \citetalias{Battistini2015A&A...577A...9B}, the differences when considering NLTE seem to be due to the use of different NLTE corrections. We compared our results of the [Sc/H], [V/H], [Mn/H], and [Co/H] ratios for the four stars in common with \citetalias{Battistini2015A&A...577A...9B} in the lower panel of Fig.~\ref{fig:comparison_BB15_DM21}, finding similar values.

\begin{figure*}
    \begin{tabular}{cc}
        \includegraphics[width=0.99\columnwidth]{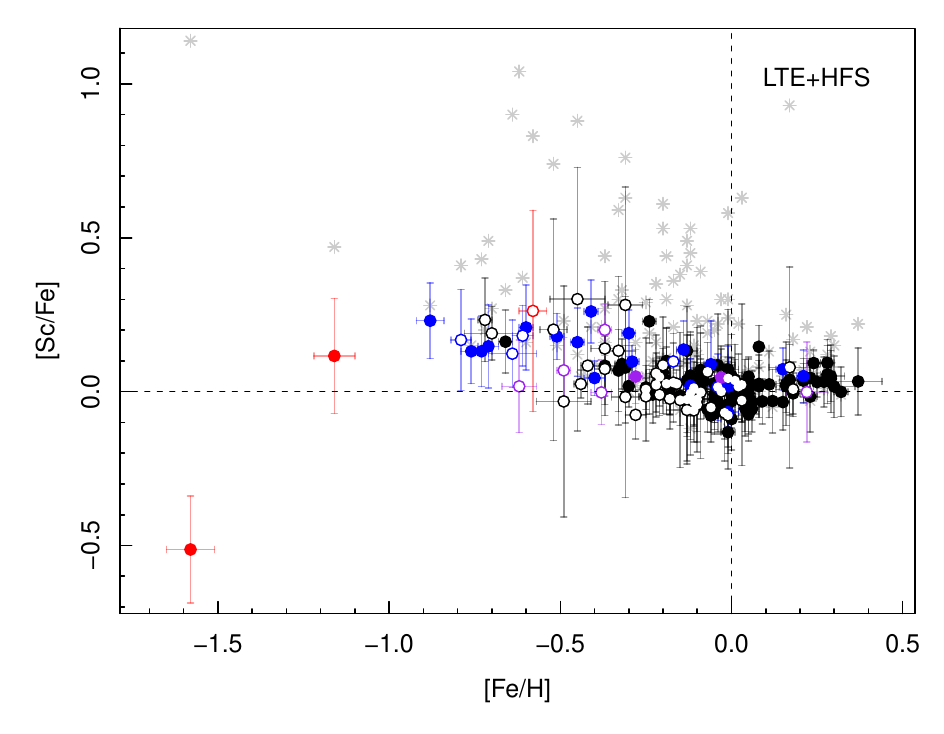} & 
        \includegraphics[width=0.99\columnwidth]{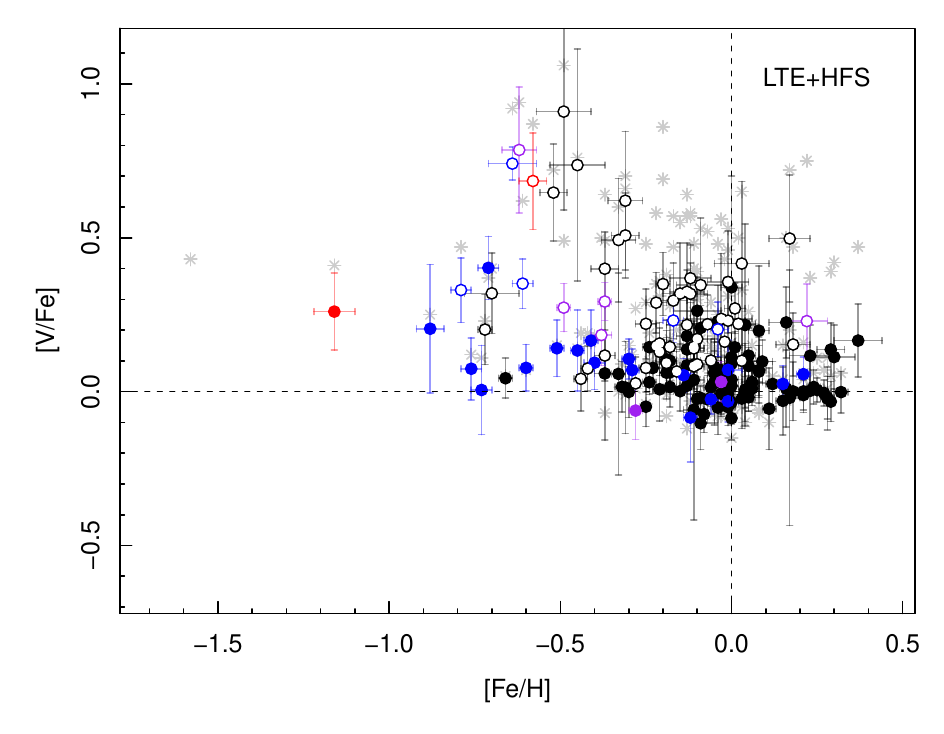} \\
        \includegraphics[width=0.99\columnwidth]{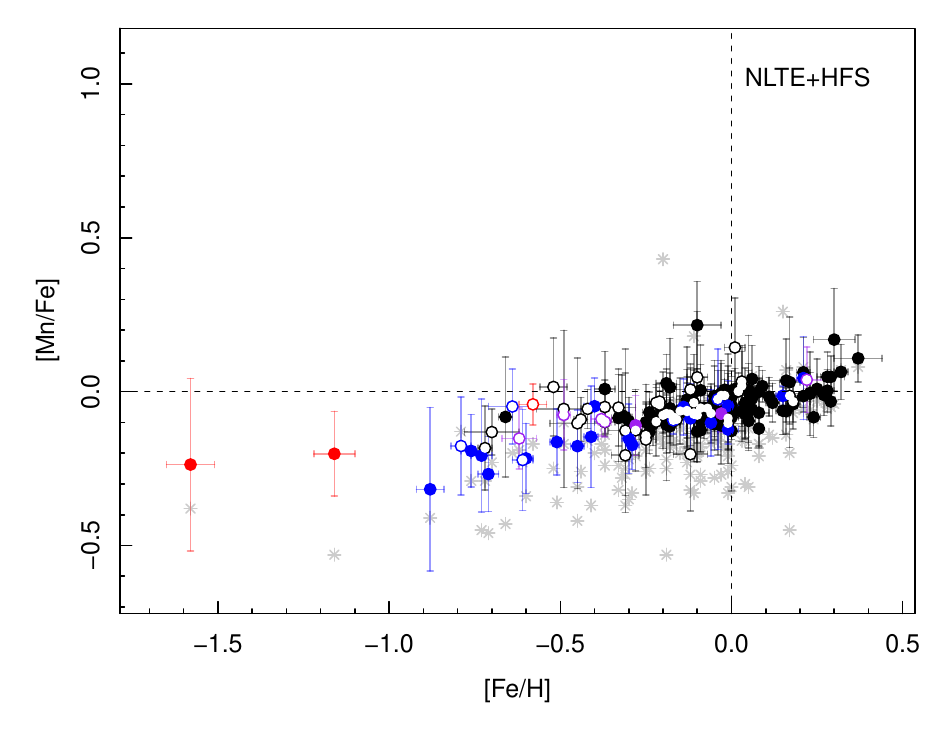} &  
        \includegraphics[width=0.99\columnwidth]{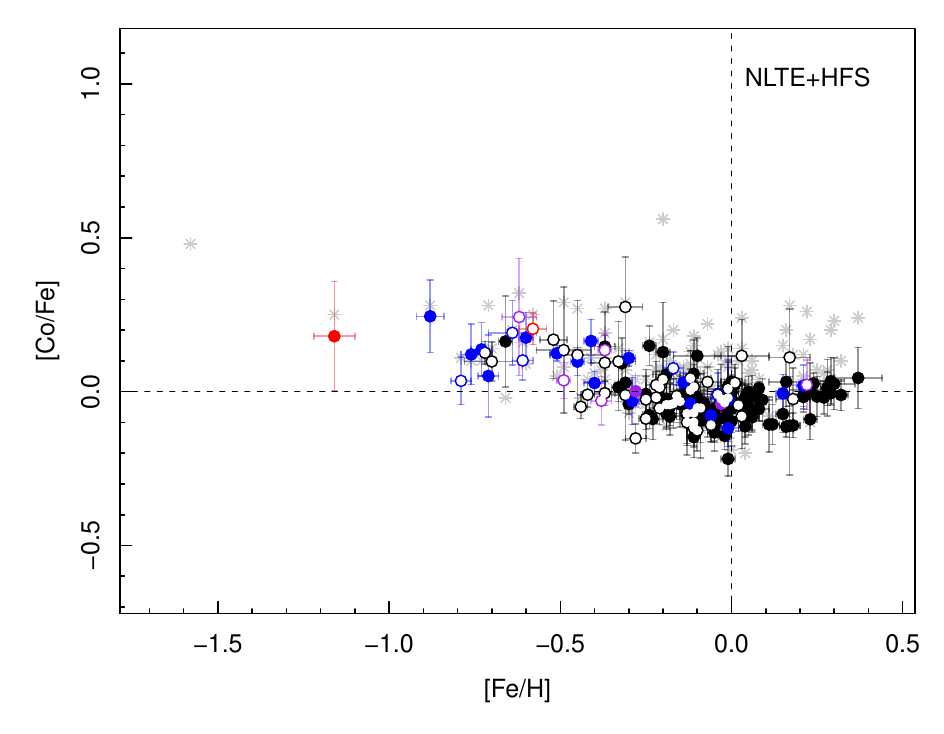} \\
    \end{tabular}

    \includegraphics[width=1.7\columnwidth]{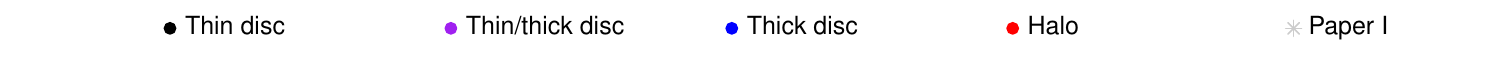} 
    
    \caption{Abundance ratios of [Sc/Fe], [V/Fe], [Mn/Fe], and [Co/Fe] versus [Fe/H], including HFS splitting and NLTE effects when necessary, in comparison with \citetalias{Montes2018MNRAS.479.1332M} (gray asterisks). Colours for different populations: black -- thin disc, blue -- thick disc, cyan -- thin/thick disc, red -- halo. Open circles: stars with $T_\text{eff}<5200$\,K; filled circles: stars with $T_\text{eff}\geq 5200$\,K.}
    \label{fig:hfsFe_paperI}
\end{figure*}

Finally we compared these results with the ones obtained in \citetalias{Montes2018MNRAS.479.1332M} without taking into account the HFS and NLTE effects in Fig.~\ref{fig:hfsFe_paperI}. We found that the outliers with $\text{[Sc/Fe]}>0.5$ disappear and the stars now follow the general Galactic trend, but the halo star BD+80 245 (G0\, IV), which presents a lower Sc abundance. Regarding the V abundances, the new results exhibit reduced scatter, with the persistent outliers attributed to lower effective temperatures, below $5200$\,K. When accounting for HFS and NLTE, we generally observed slightly higher Mn abundances and some diminished Co abundances, while maintaining the same overall trend.

\subsection{Exoplanets}

Analysing the correlations between stellar properties, such as metallicity and chemical abundances, and the occurrence of exoplanets can provide valuable insights for the selection of targets in forthcoming exoplanet surveys and for comprehending the mechanisms behind planetary formation. In particular, it has been extensively reported that the stars that host giant planets present higher metallicities than the stars without detected planets \citep[e.g.][]{Gonzalez1997MNRAS.285..403G,Fischer2005ApJ...622.1102F,Brewer2016ApJS..225...32B}. This giant planet -- stellar metallicity correlation supports the core-accretion scenario as the basis for planet formation \citep[e.g.][]{Ida2004ApJ...616..567I,Ercolano2010MNRAS.402.2735E}.
The main goal of this section is to examine potential notable distinctions between the sample of stars hosting planets and the sample of stars lacking known planets, specifically regarding metals aside from iron.

\begin{figure}
	\includegraphics[width=\columnwidth]{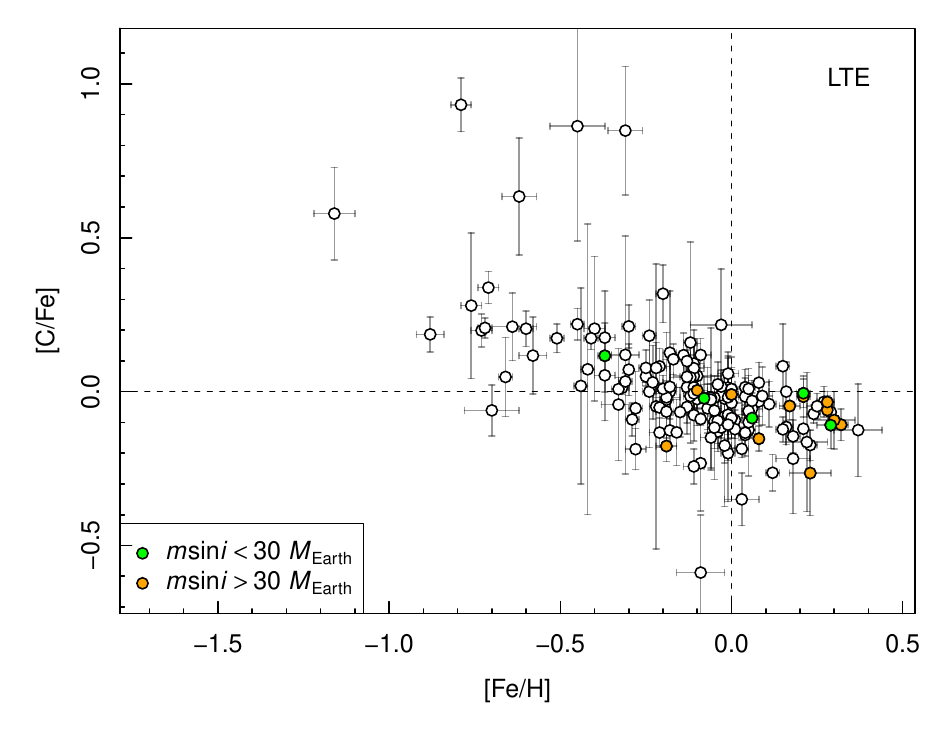}
	
	\includegraphics[width=\columnwidth]{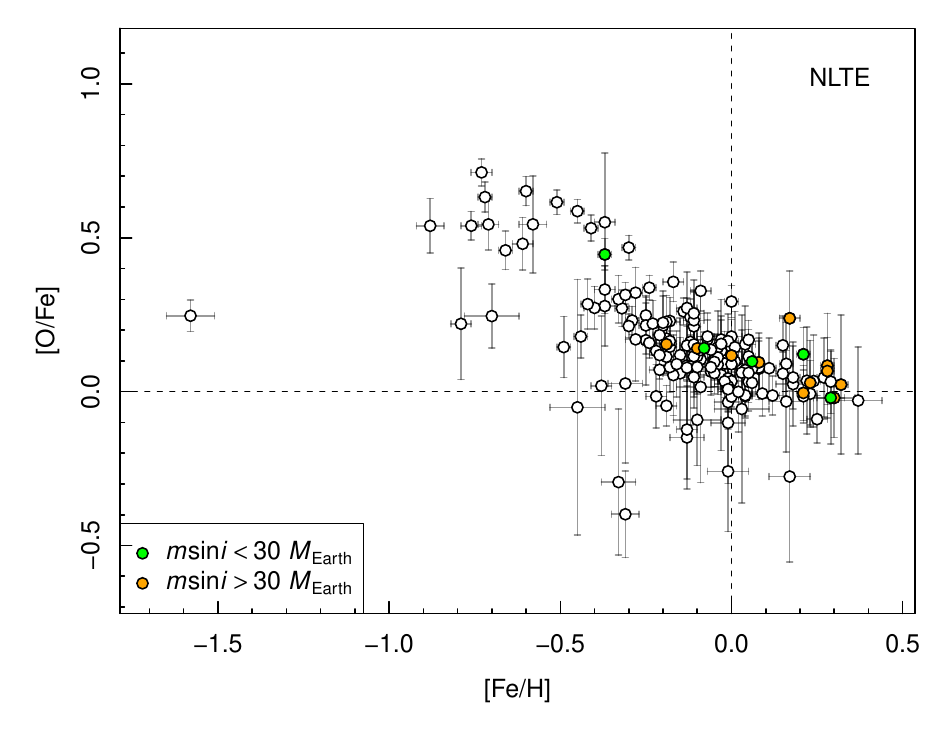}

    \includegraphics[width=\columnwidth]{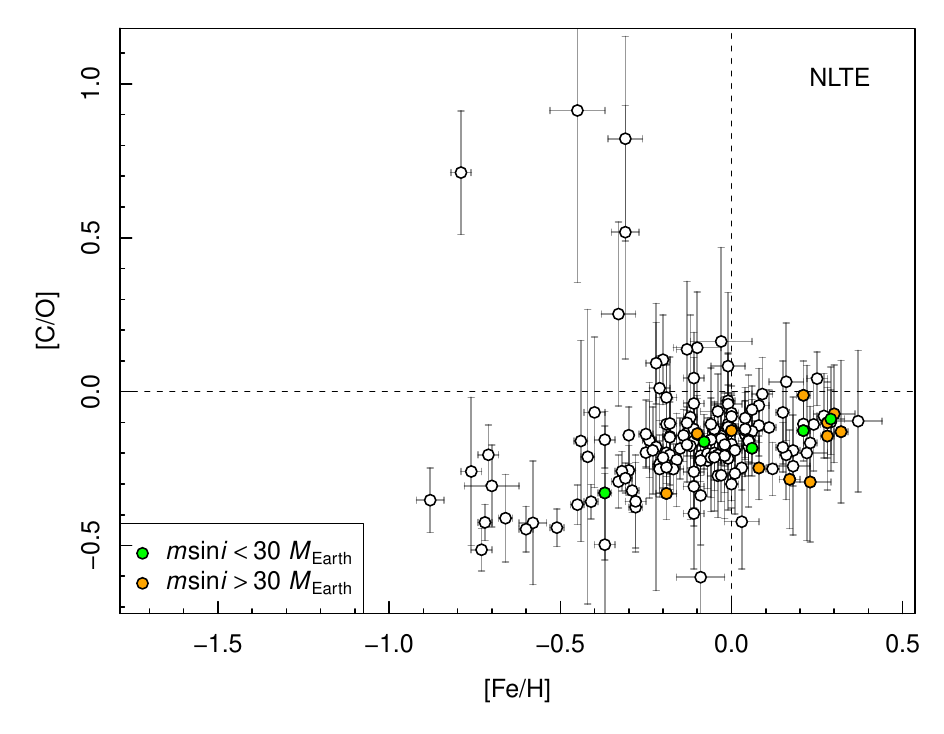}
    
    \caption{Abundance ratios of [C/Fe], [O/Fe], and [C/O] versus [Fe/H], indicating the stars with exoplanets, and colour-coded by $M\sin i$.}
    \label{fig:CFe_OFe_exoplanets}
\end{figure}

We conducted a search for confirmed exoplanets around the primary stars of our sample in the literature using The Extrasolar Planets Encyclopaedia\footnote{\url{http://exoplanet.eu}}. In \citetalias{Montes2018MNRAS.479.1332M}, 21 exoplanets in 14 planetary systems were found in the literature; now we have 24 exoplanets in 17 planetary systems. Two exoplanets were originally listed for the star HD 38529 A in \citetalias{Montes2018MNRAS.479.1332M}, but it has since been determined that one of them is a substellar companion of $23.7^{+4.1}_{-3.1}\, M_\text{Jup}$. As a result, we now recognise only one exoplanet orbiting this star. Moreover, none of the M-dwarf companions in our sample present confirmed exoplanets. We plot in Fig.~\ref{fig:CFe_OFe_exoplanets} the galactic trends for [C/Fe], [O/Fe], and [C/O], indicating the stars with confirmed exoplanets with $m\sin i$ greater or less than 30 Earth masses ($\sim 0.094\, M_\text{Jup}$), in comparison to single stars. As discussed in \citetalias{Montes2018MNRAS.479.1332M}, the exoplanet host stars tend to be metal-rich. Given the small number of exoplanet host stars in our sample, we do not see any difference between the stars harboring exoplanets with $m\sin i$ greater or less than 30 Earth masses.

\begin{figure*}
    \centering
    \includegraphics[width=2\columnwidth]{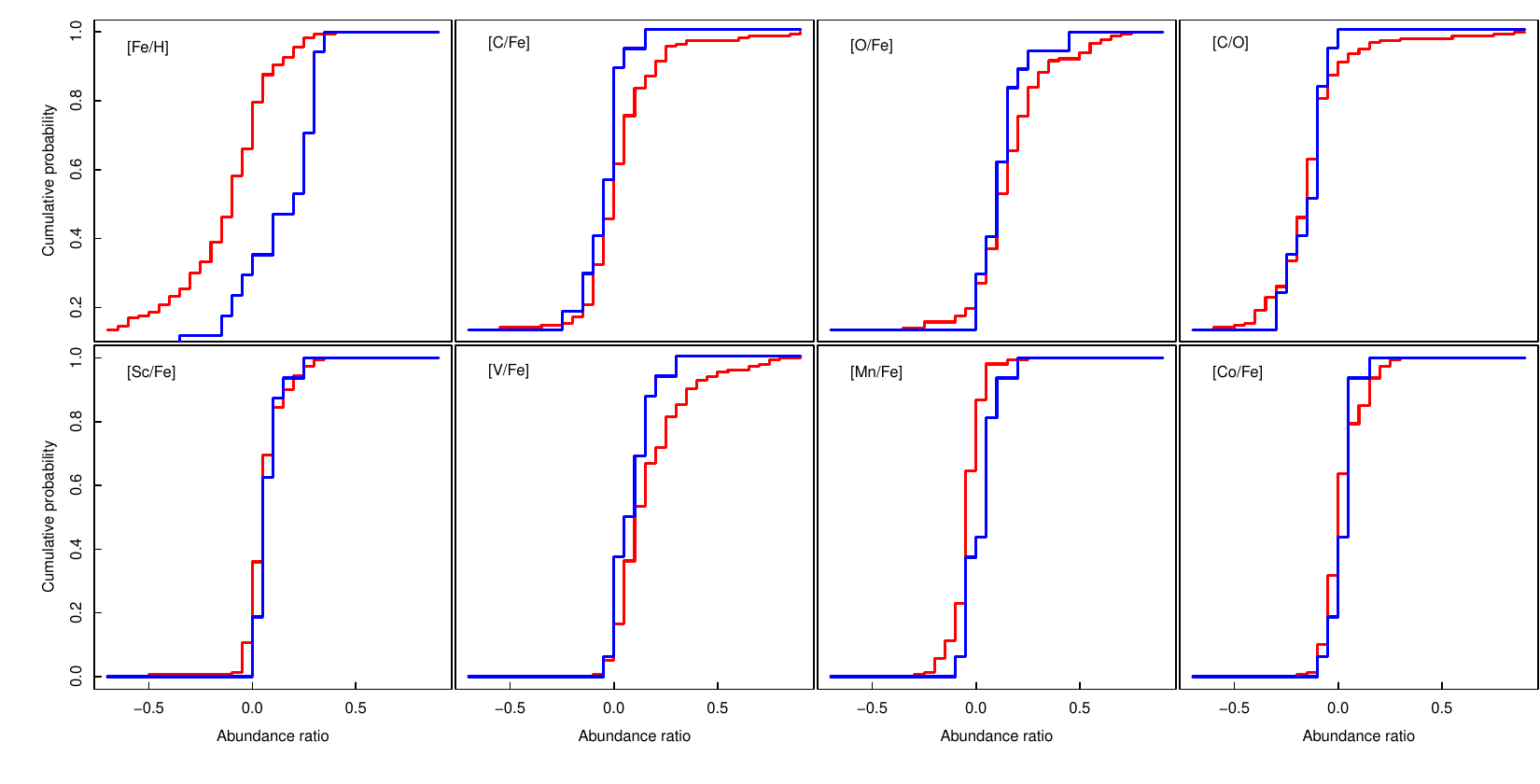}
    \caption{Normalised cumulative distribution function of the abundance ratios of several elements of the stars in our sample with (blue) and without (red) detected exoplanets.}
    \label{fig:cumulativeCO}
\end{figure*}

In \citetalias{Montes2018MNRAS.479.1332M} we performed a Kolmogorov-Smirnoff (K-S) test to evaluate the disparity between the stars in our sample with and without planets, independently of their mass, safely concluding that both metallicity distributions were significantly different. Now we repeated the K-S test for the [Fe/H], since the number of known exoplanet has increased, but we obtain the same results. The probability that both samples follow the same distribution is just of $0.0004$. We compile the K-S statistics and $p$-values in Table~\ref{tab:ks_test}, and plot the corresponding cumulative distribution function in Fig.~\ref{fig:cumulativeCO}.
The $p$-value represents the significance level of the K-S test, i.e. the probability that the two groups belong to the same population.

Preceding studies indicated that planet hosts and single stars seem to be evenly distributed for solar metallicity, and the presence of planets does not lead to differences in the carbon abundances. However, for sub-solar metallicities, planet hosts present an enhancement of C \citep[see][]{Nissen2014A&A...568A..25N,Delgado-Mena2021A&A...655A..99D}. 
We applied the K-S test for the [C/Fe], [O/Fe], and [C/O] ratios, finding that only for the [C/Fe] ratio, the distributions of host and single stars are different, with only a probability of $0.0242$ that both populations have the same [C/Fe] distribution. These results are in agreement with \citet{daSilva2015A&A...580A..24D} and \citet{Delgado-Mena2021A&A...655A..99D}, who showed a significant difference in carbon abundance between host and singles stars, but no difference, or not so strong, in the case of oxygen.

Repeating the analysis for the [X/Fe] ratios of the odd-Z iron-peak elements, we obtained that only for the [Mn/Fe] ratio, both distributions are significantly different, with just a probability of $0.0023$ for the planet hosts and single stars of coming from the same distribution. Previous studies reported overabundances of [Mn/Fe] of planet hosts in comparison to single stars \citep{Zhao2002AJ....124.2224Z,Bodaghee2003A&A...404..715B,Kang2011ApJ...736...87K,daSilva2015A&A...580A..24D}. Nevertheless, \citet{Adibekyan2012A&A...543A..89A} did not find that diference for the [Mn/Fe] ratio.

Our findings might indicate that other species than iron also play an important role in the planet formation mechanisms; for example, carbon is a pivotal element in the formation of ices within a protoplanetary disc. The overabundance of Mn could be explained by its lower condensation temperature in comparison to the other investigated elements. Notwithstanding, these results should be revised in the future with a larger sample of exoplanet hosts.

\begin{table}
    \centering
    \caption{Kolmogorov-Smirnoff test results}
    \begin{tabular}{lcc} \hline
    \noalign{\smallskip}
     Abundance  & $D$ & $p$--value \\
    \noalign{\smallskip}
    \hline
    \noalign{\smallskip}
       {[Fe/H]} & $0.5228$ & $0.0004$ \\
       {[C/Fe]} & $0.3781$ & $0.0242$ \\
       {[O/Fe]} & $0.2557$ & $0.2496$ \\
       {[C/O]} & $0.1452$ & $0.8763$ \\
       {[Sc/Fe]} & $0.2966$ & $0.1545$ \\
       {[V/Fe]} & $0.2785$ & $0.2098$ \\
       {[Mn/Fe]} & $0.4635$ & $0.0023$ \\
       {[Co/Fe]} & $0.2313$ & $0.3727$ \\
    \noalign{\smallskip}
    \hline
    \end{tabular}
    \label{tab:ks_test}
\end{table}

\section{Conclusions}
\label{Sect_conclusions}

In this paper, we extended the analysis of the chemical abundances of FGK stars with M-dwarf companions by deriving abundances of C and O, and updating the abundances of Sc, V, Mn, and Co taking into account hyperfine structure effects and correcting for non-local thermodynamic equilibrium. For that, we made use of the equivalent width method and high-resolution spectra. We compared our results with previous abundance determinations of F-, G-, and K-dwarfs found in the literature, finding a good agreement in the Galactic trends of the investigated abundances. As in previous studies, we noticed that some of the obtained abundances are not reliable for stars with $T_\text{eff}<5200$\,K, especially C, O, and V. Furthermore, we analysed the abundances of stars with and without known exoplanets and found that only for C and Mn abundances, these two populations show statistical difference.


The findings from this ongoing series devoted to investigate the metallicity and abundances of M-type dwarf stars through the use of wide physical binary systems will be beneficial for forthcoming research on the abundance of M dwarfs, which has been an expanding area of study in recent years. For instance,  \cite{Abia2020A&A...642A.227A} focused on the abundance of Rb, Zr and Sr of M dwarfs, while \cite{Shan2021A&A...654A.118S} examined V abundance, and Tabernero et al. (in prep.) explored the rock-forming elements Mg and Si. All three studies utilised CARMENES data. Upcoming papers in this series will focus on the calibration of spectral indices from low-resolution spectra of M-dwarf companions to match the chemical composition of their primary stars \citep[see][]{AlonsoFloriano2015A&A...577A.128A}. Additionally, we will use CARMENES high-resolution spectra to derive stellar atmospheric parameters and abundances of the M-dwarf companions through spectral synthesis. These abundances will then be compared against those presented here and \citetalias{Montes2018MNRAS.479.1332M} for the primary stars. By providing accurate determinations of chemical abundances in M dwarfs, this series of papers will prove invaluable for future research.

\section*{Acknowledgements}

We thank the anonymous referee for their comments and suggestions, which improved our manuscript.
This work has made use of the NASA's Astrophysics Data System; the Washington Double Star catalogue maintained at the U.S. Naval Observatory; the VALD database, operated at Uppsala University, the Institute of Astronomy RAS in Moscow, and the University of Vienna; the portal \texttt{exoplanet.eu} of The Extrasolar Planets Encyclopaedia; and data from the European Space Agency (ESA) mission {\it Gaia}. We acknowledge financial support from the Universidad Complutense de Madrid and the Agencia Estatal de Investigaci\'on (AEI/10.13039/501100011033) of the Ministerio de Ciencia e Innovaci\'on and the ERDF ``A way of making Europe'' through projects PID2019-109522GB-C5[1,4] and PID2022-137241NB-C4[2,4].

\section*{Data Availability}

The data in Table A1 are available via CDS.




\bibliographystyle{mnras}
\bibliography{example} 




\appendix


\section{Long tables}

Here we provide the tables with the abundances for the whole sample, and the HFS components for the Sc, V, Mn, and Co lines.

\begin{landscape}
\begin{table}
    \centering

    \scriptsize

    \caption{Abundances [X/H] with respect to the Sun for C, O, Sc, V, Mn, and Co.}
\begin{tabular}{llccccccccccccl} \hline
    
    \noalign{\smallskip}
         WDS & Primary & RA & DEC & Spectral & [C/H] & [O/H]$^\text{LTE}$ & [O/H]$^\text{NLTE}$ & [Sc/H] & [V/H] & [Mn/H]$^\text{LTE}$ & [Mn/H]$^\text{NLTE}$ & [Co/H]$^\text{LTE}$ & [Co/H]$^\text{NLTE}$ & Notes$^a$\\
           &  & (J2000) & (J2000) & type & (dex) & (dex) & (dex) & (dex) & (dex) & (dex) & (dex) & (dex) & (dex) \\ 
    \noalign{\smallskip}
    \hline
    \noalign{\smallskip}

    00153+5304 & G 217-41 & 00:15:14.8 & +53:04:27 & K3\,V & $-0.32\pm 0.09$ & $0.06\pm0.14$ & $0.10\pm 0.13$ & $0.00\pm 0.07$ & $0.13\pm 0.11$ & $0.07\pm0.10$ & $0.06\pm 0.10$ & $0.02\pm0.09$ & $-0.05\pm 0.11$ \\
    00385+4300 & BD+42 126 & 00:38:29.2 & +43:00:00 & G5\,V & $-0.76\pm0.08$ & $-0.52\pm0.11$ & $-0.45\pm0.10$ & $-0.51\pm0.09$ & $-0.38\pm 0.13$ & $-0.83\pm0.08$ & $-0.83\pm 0.07$ & $-0.56\pm0.06$ & $-0.60\pm0.06$ \\
    00452+0015 & HD 4271 Aa,Ab & 00:45:11.0 & +00:15:12 & F8\,V & $0.11\pm0.06$ & $0.21\pm0.12$ & $0.15\pm0.10$ & $0.10\pm0.10$ & $0.28\pm 0.21$ & $0.00\pm0.11$ & $0.01\pm 0.11$ & $0.01\pm0.05$ & $0.02\pm0.04$ \\
    00467-0426 & HD 4449 & 00:46:40.5 & $-$04:25:37 & G5\,V & $-0.05\pm0.10$ & $-0.23\pm0.17$ & $-0.19\pm0.15$ & $-0.09\pm0.20$ & $0.16\pm 0.11$ & $0.13\pm0.14$ & $0.12\pm 0.14$ & $0.04\pm0.06$ & $0.02\pm0.06$ \\
    00491+5749 & Achird Aa,Ab & 00:49:06.2 & +57:48:55 & G0\,V & $-0.20\pm0.02$ & $-0.01\pm0.05$ & $0.00\pm0.04$ & $-0.24\pm0.07$ & $-0.30\pm 0.06$ & $-0.36\pm0.07$ & $-0.35\pm 0.07$ & $-0.30\pm0.06$ & $-0.26\pm0.06$ \\
    01055+1523 & HD 6440 A & 01:05:29.9 & +15:23:24 & K3.5\,V & $\cdots$ & $-0.14\pm0.16$ & $-0.08\pm0.15$ & $-0.24\pm0.17$ & $-0.03\pm 0.11$ & $-0.35\pm0.18$ & $-0.41\pm 0.18$ & $-0.18\pm0.12$ & $-0.28\pm0.10$ \\
\hline
    \end{tabular}
    \label{tab:abundances_apendix}
\end{table}
\end{landscape}

\begin{landscape}
\begin{table}
    \centering

    \scriptsize
    
    \contcaption{}
    \begin{tabular}{llccccccccccccl} \hline
    \noalign{\smallskip}
         WDS & Primary & RA & DEC & Spectral & [C/H] & [O/H]$^\text{LTE}$ & [O/H]$^\text{NLTE}$ & [Sc/H] & [V/H] & [Mn/H]$^\text{LTE}$ & [Mn/H]$^\text{NLTE}$ & [Co/H]$^\text{LTE}$ & [Co/H]$^\text{NLTE}$ & Notes$^a$\\
           &  & (J2000) & (J2000) & type & (dex) & (dex) & (dex) & (dex) & (dex) & (dex) & (dex) & (dex) & (dex) \\ 
    \noalign{\smallskip}
    \hline
    \noalign{\smallskip}
    
    05466+0110 & HD 38529 A & 05:46:34.9 & +01:10:05 & G4\,IV & $0.21\pm0.05$ & $0.44\pm0.27$ & $0.34\pm0.23$ & $0.32\pm0.08$ & $0.32\pm0.07$ & $0.37\pm0.08$ & $0.38\pm0.09$ & $0.31\pm0.05$ & $0.31\pm0.05$ \\
    05584-0439 & HD 40397 A & 05:58:21.5 & $-$04:39:02 & G7\,V & $-0.02\pm0.07$ & $0.09\pm0.05$ & $0.12\pm0.04$ & $0.00\pm0.09$ & $-0.09\pm0.05$ & $-0.19\pm0.09$ & $-0.19\pm0.09$ & $-0.10\pm0.03$ & $-0.11\pm0.03$ \\
    06066+0431 & Ross 413 & 06:06:30.0 & +04:30:41 & K4\,V & $-0.46\pm0.13$ & $-0.11\pm0.17$ & $-0.04\pm0.16$ & $-0.32\pm0.33$ & $0.10\pm0.16$ & $-0.57\pm0.08$ & $-0.62\pm0.07$ & $-0.30\pm0.06$ & $-0.38\pm0.05$ \\
    06173+0506 & HD 43587 & 06:17:16.1 & +05:06:00 & F9\,V & $-0.01\pm0.02$ & $0.16\pm0.09$ & $0.14\pm0.08$ & $-0.02\pm0.08$ & $-0.04\pm0.07$ & $-0.05\pm0.09$ & $-0.04\pm0.10$ & $-0.09\pm0.04$ & $-0.06\pm0.04$ \\
    06314-0134 & HD 291763 & 06:31:23.1 & $-$01:34:14 & K2\,V & $\cdots$& $-0.19\pm0.09$ & $-0.13\pm0.09$ & $-0.43\pm0.10$ & $-0.26\pm0.08$ & $-0.79\pm0.15$ & $-0.83\pm0.16$ & $-0.47\pm0.06$ & $-0.51\pm0.06$ \\
    06319+0039 & HD 291725 & 06:31:51.4 & +00:38:59 & G7\,V & $-0.38\pm0.05$ & $-0.09\pm0.05$ & $-0.06\pm0.05$ & $-0.19\pm0.05$ & $-0.22\pm0.07$ & $-0.47\pm0.07$ & $-0.46\pm0.08$ & $-0.29\pm0.09$ & $-0.32\pm0.08$ \\
\hline
    \end{tabular}
\end{table}
\end{landscape}

\begin{landscape}
\begin{table}
    \centering

    \scriptsize
    
    \contcaption{}
    \begin{tabular}{llccccccccccccl} \hline
    \noalign{\smallskip}
         WDS & Primary & RA & DEC & Spectral & [C/H] & [O/H]$^\text{LTE}$ & [O/H]$^\text{NLTE}$ & [Sc/H] & [V/H] & [Mn/H]$^\text{LTE}$ & [Mn/H]$^\text{NLTE}$ & [Co/H]$^\text{LTE}$ & [Co/H]$^\text{NLTE}$ & Notes$^a$\\
           &  & (J2000) & (J2000) & type & (dex) & (dex) & (dex) & (dex) & (dex) & (dex) & (dex) & (dex) & (dex) \\ 
    \noalign{\smallskip}
    \hline
    \noalign{\smallskip}

    12049+1729 & HD 104923 & 12:04:57.0 & +17:28:36 & K0\,V & $-0.18\pm0.11$ & $-0.08\pm0.06$ & $-0.03\pm0.05$ & $-0.20\pm0.06$ & $-0.12\pm0.06$ & $-0.22\pm0.10$ & $-0.23\pm0.11$ & $-0.23\pm0.06$ & $-0.26\pm0.06$ \\
    12051+1933 & BD+20 2678 A & 12:05:07.0 & +19:33:16 & G5\,V & $-0.09\pm0.10$ & $0.09\pm0.06$ & $0.11\pm0.05$ & $-0.13\pm0.07$ & $-0.10\pm0.08$ & $-0.15\pm0.05$ & $-0.15\pm0.05$ & $-0.14\pm0.07$ & $-0.13\pm0.08$ \\
    12069+0548 & HD 105219 & 12:06:56.5 & +05:48:12 & K0\,V & $-0.18\pm0.19$ & $-0.11\pm0.06$ & $-0.08\pm0.05$ & $-0.17\pm0.04$ & $-0.13\pm0.05$ & $-0.25\pm0.06$ & $-0.26\pm0.07$ & $-0.20\pm0.05$ & $-0.21\pm0.06$ \\
    12089+2147 & BD+22 2442 & 12:08:54.7 & +21:47:19 & G2\,V+ & $-0.37\pm0.05$ & $-0.22\pm0.10$ & $-0.17\pm0.08$ & $-0.56\pm0.13$ & $-0.31\pm0.10$ & $-0.97\pm0.11$ & $-0.98\pm0.12$ & $-0.71\pm0.10$ & $-0.66\pm0.13$ \\
    12372+3545 & BD+36 2288 & 12:37:13.7 & +35:44:46 & G5\,V & $-0.15\pm0.09$ & $0.05\pm0.10$ & $0.05\pm0.09$ & $-0.06\pm0.09$ & $-0.02\pm0.13$ & $-0.09\pm0.10$ & $-0.08\pm0.11$ & $-0.11\pm0.07$ & $-0.14\pm0.07$ \\
    12406+4017 & HD 110279 & 12:40:37.4 & +40:17:17 & F8\,V & $-0.09\pm0.07$ & $0.03\pm0.10$ & $0.03\pm0.08$ & $0.05\pm0.06$ & $0.11\pm0.08$ & $-0.13\pm0.17$ & $-0.13\pm0.19$ & $-0.07\pm0.06$ & $-0.06\pm0.06$ \\
\hline
    \end{tabular}
\end{table}
\end{landscape}

\begin{landscape}
\begin{table}
    \centering

    \scriptsize
    
    \contcaption{}
    \begin{tabular}{llccccccccccccl} \hline
    \noalign{\smallskip}
         WDS & Primary & RA & DEC & Spectral & [C/H] & [O/H]$^\text{LTE}$ & [O/H]$^\text{NLTE}$ & [Sc/H] & [V/H] & [Mn/H]$^\text{LTE}$ & [Mn/H]$^\text{NLTE}$ & [Co/H]$^\text{LTE}$ & [Co/H]$^\text{NLTE}$ & Notes$^a$\\
           &  & (J2000) & (J2000) & type & (dex) & (dex) & (dex) & (dex) & (dex) & (dex) & (dex) & (dex) & (dex) \\ 

    \noalign{\smallskip}
    \hline
    \noalign{\smallskip}
    
    17050-0504 & HD 154363 A & 17:05:03.4 & $-$05:03:59 & K5\,V & $0.01\pm0.19$ & $\cdots$ & $\cdots$& $-0.60\pm0.15$ & $0.17\pm0.20$ & $-0.69\pm0.16$ & $-0.77\pm0.10$ & $-0.37\pm0.21$ & $-0.38\pm0.19$ \\
    17178+5227 & HD 156985 & 17:17:50.4 & +52:26:50 & K2\,V & $-0.14\pm0.08$ & $-0.29\pm0.11$ & $-0.24\pm0.10$ & $-0.16\pm0.13$ & $0.07\pm0.10$ & $-0.27\pm0.12$ & $-0.32\pm0.11$ & $-0.16\pm0.08$ & $-0.20\pm0.08$ \\
    17272+4213 & HD 158415 & 17:27:13.9 & +42:13:05 & G5\,V & $-0.14\pm0.06$ & $0.11\pm0.07$ & $0.11\pm0.06$ & $0.09\pm0.10$ & $0.14\pm0.07$ & $0.08\pm0.06$ & $0.08\pm0.06$ & $0.03\pm0.05$ & $0.01\pm0.07$ \\
    17411+7225 & HD 161897 & 17:41:06.7 & +72:25:13 & G6\,V & $-0.09\pm0.04$ & $-0.04\pm0.04$ & $-0.02\pm0.04$ & $-0.01\pm0.08$ & $0.02\pm0.05$ & $-0.07\pm0.05$ & $-0.08\pm0.05$ & $-0.02\pm0.09$ & $-0.05\pm0.08$ \\
    17428+1646 & BD+16 3263 & 17:42:50.5 & +16:45:54 & K0\,V & $-0.20\pm0.12$ & $-0.05\pm0.09$ & $-0.01\pm0.08$ & $-0.24\pm0.09$ & $-0.15\pm0.08$ & $-0.33\pm0.06$ & $-0.34\pm0.07$ & $-0.30\pm0.07$ & $-0.32\pm0.07$ \\
    17465+2743 & $\mu^{01}$ Her A & 17:46:27.5 & +27:43:14 & G5\,IV+ & $0.24\pm0.05$ & $0.37\pm0.15$ & $0.32\pm0.13$ & $0.30\pm0.08$ & $0.27\pm0.04$ & $0.26\pm0.06$ & $0.26\pm0.07$ & $0.27\pm0.04$ & $0.25\pm0.05$ \\
\hline
    \end{tabular}

    \footnotesize{$^a$ SB2: Double-line spectroscopic binary; Hot: Spectral type $\leq$ F6\,V; Cool: Spectral type $\geq$ K4\,V; Fast: $v\sin i\geq10$\, km\,s$^{-1}$}
    
\end{table}
\end{landscape}


\begin{landscape}
\begin{table}
    \centering

    \scriptsize
    
    \caption{HFS components for the Sc lines.}
    \begin{tabular}{cccccccccccccccc} \hline
    \noalign{\smallskip}
         \multicolumn{2}{c}{\ion{Sc}{I} 5520.519} & \multicolumn{2}{c}{\ion{Sc}{I} 5671.860} & \multicolumn{2}{c}{\ion{Sc}{II} 5526.789} & \multicolumn{2}{c}{\ion{Sc}{II} 5657.886} & \multicolumn{2}{c}{\ion{Sc}{II} 5667.135} & \multicolumn{2}{c}{\ion{Sc}{II} 5684.190} & \multicolumn{2}{c}{\ion{Sc}{II} 6245.621} & \multicolumn{2}{c}{\ion{Sc}{II} 6320.832} \\ \hline
         $\lambda$ (\AA) & $\log(gf)$ & $\lambda$ (\AA) & $\log(gf)$ & $\lambda$ (\AA) & $\log(gf)$ & $\lambda$ (\AA) & $\log(gf)$ & $\lambda$ (\AA) & $\log(gf)$ & $\lambda$ (\AA) & $\log(gf)$ & $\lambda$ (\AA) & $\log(gf)$ & $\lambda$ (\AA) & $\log(gf)$ \\

    \noalign{\smallskip}
    \hline
    \noalign{\smallskip}

    5520.482 & $-0.339$ & 5671.773 & $-0.926$ & 5526.770 & $-2.805$ & 5657.886 & $-1.166$ & 5667.135 & $-1.804$ & 5684.190 & $-1.553$ & 6245.621 & $-1.624$ & 6320.832 & $-2.410$ \\
    5520.494 & $-0.339$ & 5671.778 & $-0.808$ & 5526.775 & $-2.402$ & 5657.887 & $-1.736$ & 5667.141 & $-2.000$ & 5684.192 & $-2.042$ & 6245.629 & $-2.364$ & 6320.839 & $-2.606$ \\
    5520.508 & $-0.339$ & 5671.779 & $-1.378$ & 5526.779 & $-1.592$ & 5657.892 & $-1.736$ & 5667.148 & $-2.000$ & 5684.193 & $-2.741$ & 6245.631 & $-1.795$ & 6320.850 & $-2.606$ \\
    5520.519 & $-0.339$ & 5671.785 & $-0.690$ & 5526.779 & $-2.190$ & 5657.894 & $-1.564$ & 5667.153 & $-3.185$ & 5684.204 & $-1.846$ & 6245.636 & $-3.364$ & 6320.857 & $-3.791$ \\
             &          & 5671.787 & $-1.167$ & 5526.782 & $-1.387$ & 5657.895 & $-1.578$ & 5667.157 & $-2.004$ & 5684.206 & $-1.926$ & 6245.638 & $-2.181$ & 6320.862 & $-2.610$ \\
             &          & 5671.788 & $-2.378$ & 5526.783 & $-2.080$ & 5657.899 & $-1.578$ & 5667.162 & $-2.004$ & 5684.207 & $-2.301$ & 6245.640 & $-2.002$ & 6320.871 & $-2.610$ \\
\hline
    \end{tabular}

    
    \label{tab:HFS_Sc}
\end{table}

\begin{table}
    \centering

    \scriptsize
    
    \caption{HFS components for the V lines.}
    \begin{tabular}{cccccccccccccc} \hline
    \noalign{\smallskip}
         \multicolumn{2}{c}{\ion{V}{I} 5670.831} & \multicolumn{2}{c}{\ion{V}{I} 5737.027} & \multicolumn{2}{c}{\ion{V}{I} 6081.461} & \multicolumn{2}{c}{\ion{V}{I} 6224.520} & \multicolumn{2}{c}{\ion{V}{I} 6251.802} & \multicolumn{2}{c}{\ion{V}{I} 6274.659} & \multicolumn{2}{c}{\ion{V}{I} 6285.162} \\ \hline
         $\lambda$ (\AA) & $\log(gf)$ & $\lambda$ (\AA) & $\log(gf)$ & $\lambda$ (\AA) & $\log(gf)$ & $\lambda$ (\AA) & $\log(gf)$ & $\lambda$ (\AA) & $\log(gf)$ & $\lambda$ (\AA) & $\log(gf)$ & $\lambda$ (\AA) & $\log(gf)$ \\

    \noalign{\smallskip}
    \hline
    \noalign{\smallskip}

    5670.811 & $-1.103$ & 5737.027 & $-1.386$ & 6081.418 & $-1.814$ & 6224.486 & $-3.594$ & 6251.767 & $-2.954$ & 6274.611 & $-2.962$ & 6285.108 & $-3.598$ \\
    5670.819 & $-2.111$ & 5737.037 & $-2.085$ & 6081.418 & $-1.638$ & 6224.495 & $-3.390$ & 6251.784 & $-2.750$ & 6274.632 & $-2.485$ & 6285.128 & $-3.170$ \\
    5670.826 & $-3.412$ & 5737.041 & $-2.085$ & 6081.428 & $-1.638$ & 6224.504 & $-3.325$ & 6251.799 & $-2.685$ & 6274.644 & $-2.506$ & 6285.133 & $-2.733$ \\
    5670.831 & $-1.208$ & 5737.051 & $-1.678$ & 6081.429 & $-2.610$ & 6224.513 & $-3.332$ & 6251.802 & $-2.051$ & 6274.659 & $-2.164$ & 6285.144 & $-2.920$ \\
    5670.838 & $-1.900$ & 5737.059 & $-1.904$ & 6081.429 & $-1.513$ & 6224.520 & $-2.691$ & 6251.813 & $-2.692$ & 6274.660 & $-2.485$ & 6285.147 & $-2.572$ \\
    5670.844 & $-3.014$ & 5737.063 & $-1.904$ & 6081.443 & $-1.513$ & 6224.521 & $-3.398$ & 6251.814 & $-2.238$ & 6274.682 & $-2.631$ & 6285.158 & $-2.744$ \\
\hline
    \end{tabular}

    
    \label{tab:HFS_V}
\end{table}
\end{landscape}

\begin{landscape}
\begin{table}
    \centering

    \scriptsize
    
    \caption{HFS components for the Mn lines.}
    \begin{tabular}{cccccccccccc} \hline
    \noalign{\smallskip}
         \multicolumn{2}{c}{\ion{Mn}{I} 4502.204} & \multicolumn{2}{c}{\ion{Mn}{I} 4671.647} & \multicolumn{2}{c}{\ion{Mn}{I} 4739.072} & \multicolumn{2}{c}{\ion{Mn}{I} 5377.626} & \multicolumn{2}{c}{\ion{Mn}{I} 5399.500} & \multicolumn{2}{c}{\ion{Mn}{I} 5413.697} \\ \hline
         $\lambda$ (\AA) & $\log(gf)$ & $\lambda$ (\AA) & $\log(gf)$ & $\lambda$ (\AA) & $\log(gf)$ & $\lambda$ (\AA) & $\log(gf)$ & $\lambda$ (\AA) & $\log(gf)$ & $\lambda$ (\AA) & $\log(gf)$ \\

    \noalign{\smallskip}
    \hline
    \noalign{\smallskip}
    
    4502.204 & $-0.912$ & 4671.647 & $-2.242$ & 4739.070 & $-1.635$ & 5377.564 & $-1.576$ & 5399.423 & $-1.403$ & 5413.593 & $-1.981$ \\
    4502.207 & $-1.830$ & 4671.654 & $-3.160$ & 4739.072 & $-1.158$ & 5377.576 & $-1.166$ & 5399.429 & $-1.271$ & 5413.628 & $-1.550$ \\
    4502.208 & $-3.047$ & 4671.660 & $-4.377$ & 4739.090 & $-1.533$ & 5377.576 & $-1.400$ & 5399.439 & $-1.771$ & 5413.638 & $-1.437$ \\
    4502.211 & $-1.052$ & 4671.664 & $-2.382$ & 4739.091 & $-1.707$ & 5377.582 & $-1.722$ & 5399.441 & $-1.373$ & 5413.658 & $-1.437$ \\
    4502.213 & $-1.654$ & 4671.670 & $-2.984$ & 4739.093 & $-1.635$ & 5377.594 & $-1.245$ & 5399.453 & $-2.980$ & 5413.697 & $-1.073$ \\
    4502.213 & $-2.695$ & 4671.675 & $-4.025$ & 4739.105 & $-1.665$ & 5377.596 & $-0.893$ & 5399.468 & $-1.403$ & 5413.702 & $-1.534$ \\
\hline
    \end{tabular}

    
    \label{tab:HFS_Mn}
\end{table}

\begin{table}
    \centering

    \scriptsize
    
    \caption{HFS components for the Co lines.}
    \begin{tabular}{cccccccccccccccccc} \hline
    \noalign{\smallskip}
         \multicolumn{2}{c}{\ion{Co}{I} 4594.669} & \multicolumn{2}{c}{\ion{Co}{I} 4792.854} & \multicolumn{2}{c}{\ion{Co}{I} 4813.471} & \multicolumn{2}{c}{\ion{Co}{I} 5301.017} & \multicolumn{2}{c}{\ion{Co}{I} 5342.706} & \multicolumn{2}{c}{\ion{Co}{I} 5352.020} & \multicolumn{2}{c}{\ion{Co}{I} 5359.188} & \multicolumn{2}{c}{\ion{Co}{I} 5647.207} & \multicolumn{2}{c}{\ion{Co}{I} 6814.958} \\ \hline
         $\lambda$ (\AA) & $\log(gf)$ & $\lambda$ (\AA) & $\log(gf)$ & $\lambda$ (\AA) & $\log(gf)$ & $\lambda$ (\AA) & $\log(gf)$ & $\lambda$ (\AA) & $\log(gf)$ & $\lambda$ (\AA) & $\log(gf)$ & $\lambda$ (\AA) & $\log(gf)$ & $\lambda$ (\AA) & $\log(gf)$ & $\lambda$ (\AA) & $\log(gf)$ \\

    \noalign{\smallskip}
    \hline
    \noalign{\smallskip}

    4594.582 & $-1.848$ & 4792.799 & $-2.583$ & 4813.398 & $-2.862$ & 5301.008 & $-3.345$ & 5342.693 & $-0.459$ & 5351.893 & $-3.244$ & 5359.188 & $-0.429$ & 5647.207 & $-2.127$ & 6814.889 & $-2.917$ \\
    4594.583 & $-1.568$ & 4792.814 & $-2.173$ & 4813.416 & $-2.464$ & 5301.017 & $-2.646$ & 5342.693 & $-0.542$ & 5351.925 & $-2.855$ & 5359.191 & $-0.534$ & 5647.220 & $-2.343$ & 6814.915 & $-2.803$ \\
    4594.587 & $-1.430$ & 4792.824 & $-1.469$ & 4813.432 & $-1.561$ & 5301.026 & $-3.164$ & 5342.693 & $-0.373$ & 5351.953 & $-1.791$ & 5359.194 & $-0.650$ & 5647.232 & $-2.626$ & 6814.924 & $-2.506$ \\
    4594.588 & $-2.869$ & 4792.827 & $-1.951$ & 4813.432 & $-2.260$ & 5301.034 & $-2.938$ & 5342.694 & $-0.289$ & 5351.976 & $-1.573$ & 5359.197 & $-0.779$ & 5647.238 & $-2.753$ & 6814.936 & $-2.928$ \\
    4594.590 & $-1.848$ & 4792.834 & $-1.273$ & 4813.445 & $-1.350$ & 5301.040 & $-3.133$ & 5342.696 & $-0.208$ & 5351.978 & $-2.583$ & 5359.200 & $-0.925$ & 5647.243 & $-3.065$ & 6814.943 & $-3.122$ \\
\hline
    \end{tabular}

    
    \label{tab:HFS_Co}
\end{table}
\end{landscape}



\bsp	
\label{lastpage}
\end{document}